\documentclass{article}
\usepackage{arxiv}

\usepackage[utf8]{inputenc} 
\usepackage[T1]{fontenc}    
\usepackage{hyperref}       
\usepackage{url}            
\usepackage{booktabs}       
\usepackage{amsfonts}       
\usepackage{nicefrac}       
\usepackage{microtype}      
\usepackage{lipsum}		
\usepackage{graphicx}
\usepackage{natbib}
\usepackage{doi}
\usepackage[section]{placeins}
\usepackage{float}

\title{SG-ML: Smart Grid Cyber Range Modelling Language}

\author{ \href{https://orcid.org/0000-0002-4761-1736}{\includegraphics[scale=0.06]{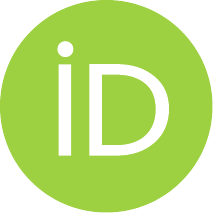}\hspace{1mm}Muhammad M. Roomi}\\
Illinois Advanced Research Center\\
Singapore \\
\texttt{roomi.s@iarcs-create.edu.sg} \\
\And
\href{https://orcid.org/0000-0002-7779-8140}{\includegraphics[scale=0.06]{orcid.pdf}\hspace{1mm}Suhail S.M. Hussain} \\
King Fahd University of Petroleum and Minerals\\
Dhahran, Saudi Arabia \\	
\texttt{s.suhail.md@gmail.com}\\
\And
\href{https://orcid.org/0000-0003-1946-1790}{\includegraphics[scale=0.06]{orcid.pdf}\hspace{1mm}Daisuke Mashima} \\
Singapore University of Technology and Design\\
Singapore \\
\texttt{daisuke\_mashima@sutd.edu.sg} \\
}

\date{}

\renewcommand{\headeright}{Technical Report}
\renewcommand{\undertitle}{Technical Report}
\renewcommand{\shorttitle}{SG-ML Modelling Language}

\begin{document} 

\maketitle
\begin{abstract}
This work provides a detailed specification of the Smart Grid Modelling Language (SG-ML), which is designed for the automated generation of smart grid cyber ranges. SG-ML is defined as a set of XML schemas that describe a smart grid’s configuration in both 
machine-readable and human-friendly ways, thereby bridging the gap between system modelling and automated deployment. Unlike prior ad-hoc approaches to cyber range design, 
SG-ML provides a unified methodology that integrates both power system and cyber network 
representations. The SG-ML model can be customized by users to meet specific requirements, such as emulating physical or cyber topologies and configuring network devices. An SG-ML Processor then parses this configured model to instantiate the cyber range environment. The modelling language leverages established standards like the IEC 61850 Substation Configuration Language (SCL) and IEC 61131 PLCopen XML to define power system topology, cyber network topology, and device configurations. This approach allows for the reuse of existing assets, reducing the effort needed to create the SG-ML model. To address gaps not covered by these standards such as attack injection parameters, scenario-specific metadata, and additional network constraints, SG-ML introduces proprietary schemas that complement standard models. Overall, SG-ML enables reproducible, scalable, and automated generation of realistic smart grid cyber ranges for research, training, and security assessment.
\end{abstract}

\keywords{Smart Grid Modelling Language (SG-ML) \and Cyber Range \and IEC 61850 Substation Configuration Language (SCL) \and XML Schemas \and Cyber-Physical Systems}

\section{Introduction}\label{sec:intro}
This paper presents a detailed specification of the Smart Grid Modelling Language (SG-ML), developed for the automated generation of smart grid cyber ranges. SG-ML is structured as a set of XML schemas that enable the description of a cyber range configuration in a way that is both machine-readable and human-interpretable, bridging the gap between system modelling and automated deployment. One of the key strengths of SG-ML is its customizability: users can tailor the model to match their specific requirements, such as the physical or cyber topology to emulate, the types and roles of networked devices, and the operational scenarios to be tested. Once a configuration is created, the SG-ML model is parsed in accordance with the defined schema and utilized by a toolchain called the SG-ML Processor, which instantiates the actual cyber range environment automatically.

Cyber ranges play an increasingly critical role in modern power system research, training, and security assessment. They provide controlled environments in which operators, researchers, and trainees can simulate realistic operational conditions and evaluate the impact of cyber-attacks without endangering real infrastructure. For power grids, the need for such environments is heightened by the growing integration of digital technologies, smart devices, and communication networks, which expand both operational capabilities and the potential attack surface. Designing a cyber range that accurately represents both the electrical and cyber domains, however, remains a significant challenge due to the complexity, heterogeneity, and scale of smart grid systems. SG-ML is conceived to address these challenges by offering a systematic and extensible way to specify cyber range configurations.

SG-ML includes XML schemas for defining three essential categories of information required to instantiate a smart grid cyber range:

\begin{enumerate}
    \item Power system topology and configuration: This includes the representation of single-line diagrams, the arrangement and operational status of power system components (e.g., circuit breakers, transformers, generators), and the load profiles associated with various nodes.
    \item Cyber network topology and configuration: This defines the connectivity between devices, network parameters such as bandwidth and latency, communication protocols, and segmentation of the network for security and operational purposes.
    \item Device configurations: This encompasses the setup of individual devices, including network addresses, communication models, and the functional roles of SCADA HMIs, PLCs, IEDs, and other embedded devices.
\end{enumerate}

To ensure interoperability and reduce modelling overhead, SG-ML leverages widely-adopted standards, specifically the IEC 61850 Substation Configuration Language (SCL) and IEC 61131 PLCopen XML. For example, IEC 61850 SCL provides schemas for defining substation single-line diagrams, cyber network topology, communication protocols, and device attributes. By integrating these existing models into SG-ML, system operators can reuse pre-existing SCL files from their substations to accelerate the cyber range modelling process, minimizing manual effort and ensuring alignment with real-world configurations.

However, studies of standard models indicate that they do not capture all the information necessary for fully specifying a cyber range. To bridge this gap, proprietary schemas were developed within SG-ML, providing supplementary metadata and configuration details required for automated range instantiation, such as scenario-specific parameters, attack injection points, and additional network constraints.

\begin{figure}[b]
\centering
\includegraphics[width=\textwidth]{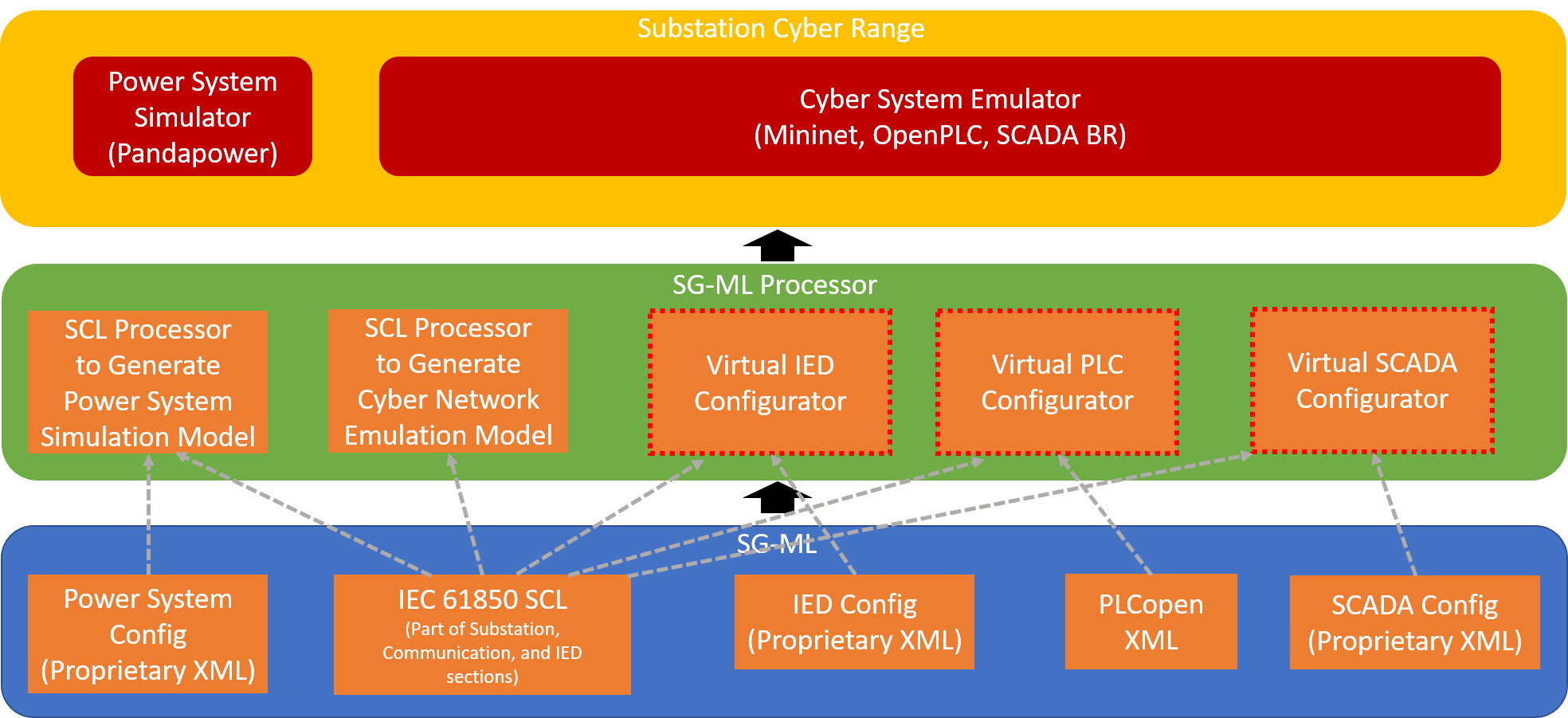}
\caption{SG-ML Overview}\label{fig:sgml}
\end{figure}

\begin{figure}[t]
\centering
{\includegraphics[width=\textwidth]{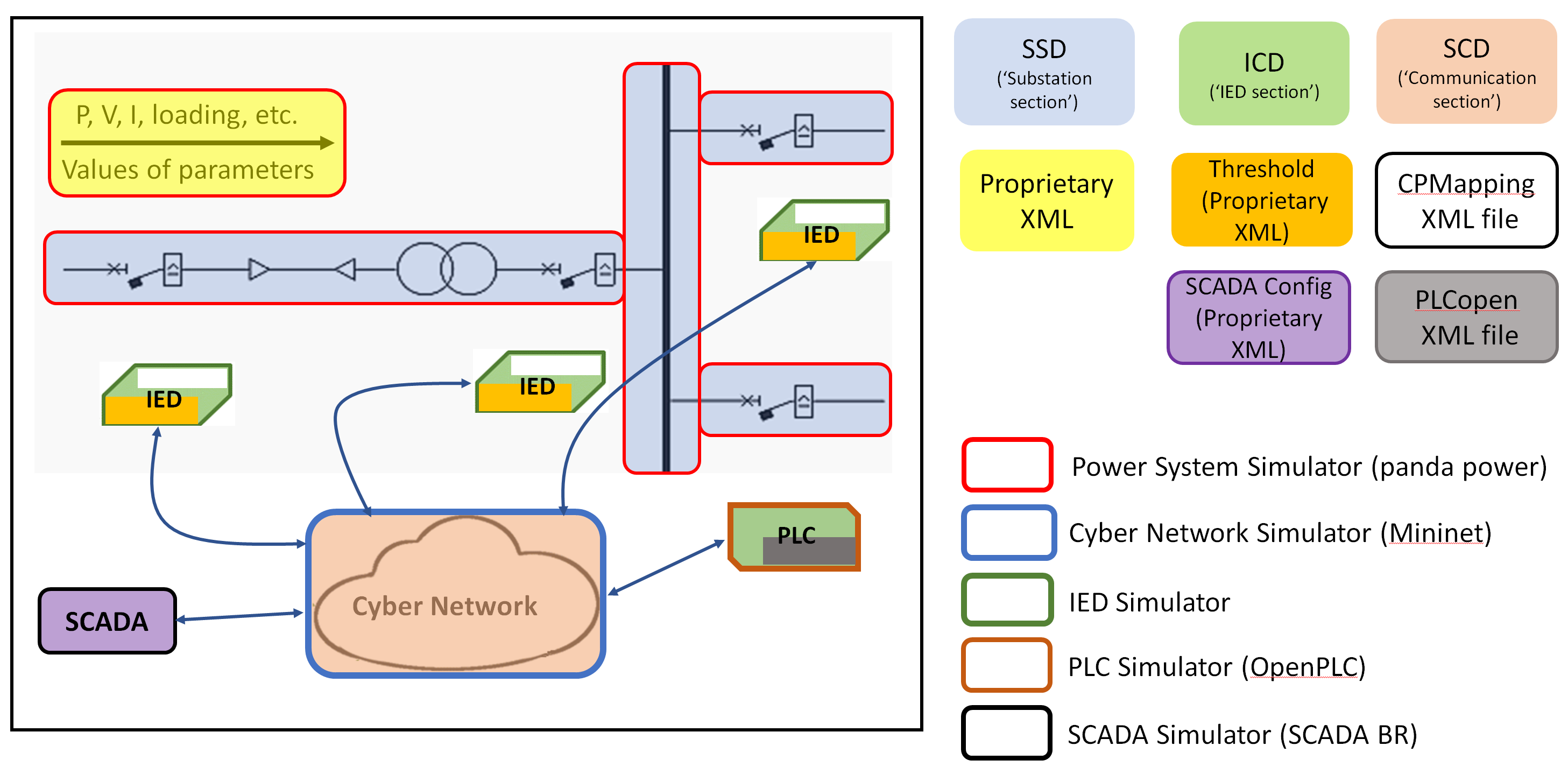}}
\caption{SG-ML modelling language}\label{sgmlframe}
\end{figure}

In summary, the SG-ML modelling language consists of multiple components and processes, as illustrated in the lower portion of Fig.~\ref{fig:sgml}. The remainder of this paper elaborates on how these components interact for the automated generation of smart grid cyber ranges, while providing a detailed overview of the standard models and proprietary extensions incorporated into the modelling language. The overall architecture and representation of the smart grid modelling language are depicted in Fig.~\ref{sgmlframe}, highlighting the interplay between physical power system models, cyber network configurations, and device-level specifications. 

The remainder of this paper is organized as follows. Section~\ref{sec:intro} presents the introduction. Section~\ref{sec:systems} describes the systems considered for modelling. Section~\ref{sec:scl} details the SCL files and the input data required for the SG-ML modelling language. Sections~\ref{sec:sgml1}–\ref{sec:sgml5} explain the application of the modelling language for automated cyber range generation. Finally, Section~\ref{sec:conclusion} concludes the paper.

\section{Systems under study} \label{sec:systems}

\subsection{Electric Power and Intelligent Control Testbed:}

\begin{figure}[h]
\centering
{\includegraphics[width=\textwidth]{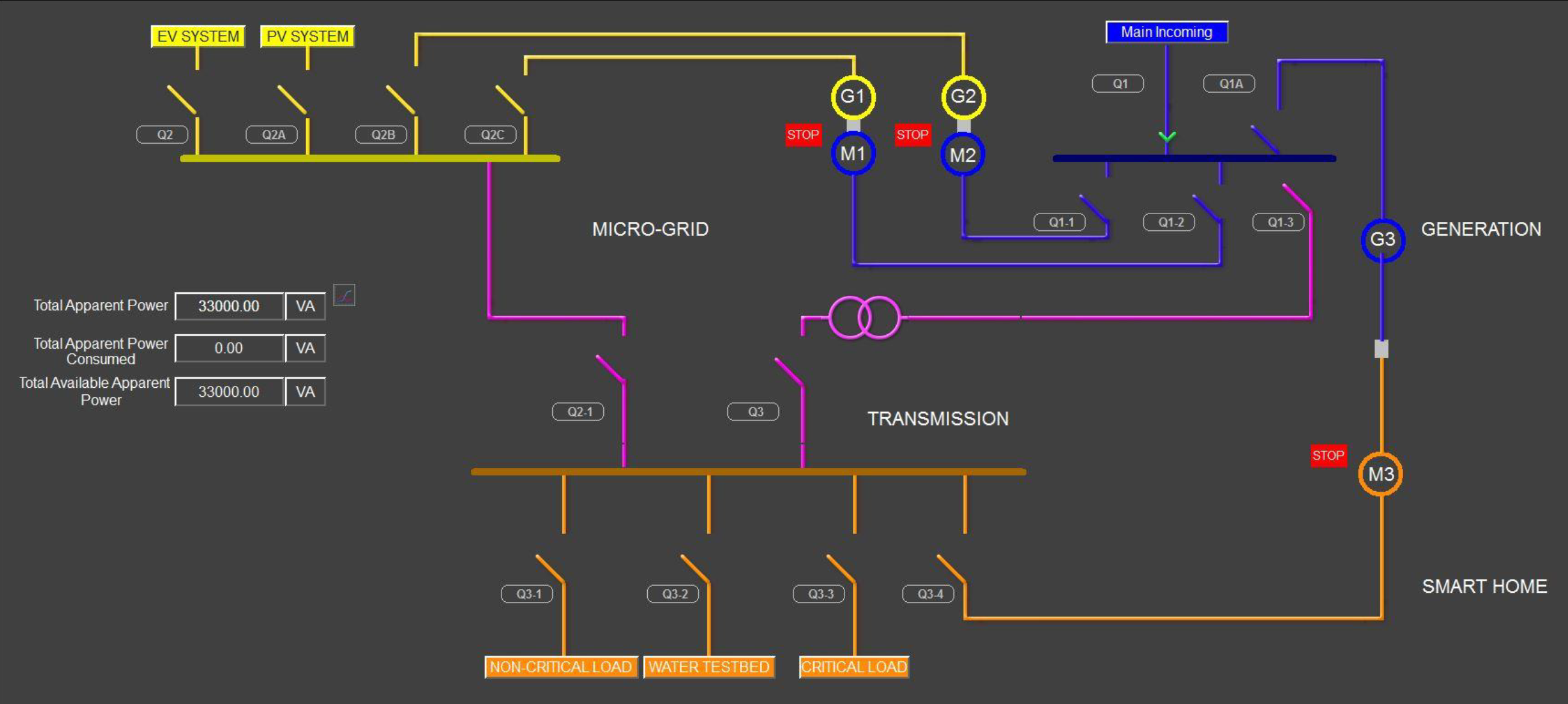}}
\caption{Overview of EPIC testbed}
\label{epic}
\end{figure}

The Electric Power and Intelligent Control (EPIC) testbed~\cite{epic}, developed at iTrust, Singapore University of Technology and Design, is a sophisticated cyber-physical platform that emulates a modern smart grid for research and training purposes. The layout of the EPIC testbed is depicted in Fig.~\ref{epic}. It combines realistic physical layers—including power generation units, transmission infrastructure, micro-grid operations, and smart homes—with cyber components such as segmented communication networks, SCADA systems, and programmable logic controllers (PLCs). The setup comprises one SCADA server, five PLC units, three generators, three variable speed drives (VSDs), twelve intelligent electronic devices (IEDs), ten network switches, and supporting equipment. Generator speed regulation is achieved through VSDs, while information exchange among IEDs, PLCs, and SCADA is conducted via the Manufacturing Message Specification (MMS) protocol, fully compliant with IEC 61850. Peer-to-peer communication among IEDs is enabled through Generic Object Oriented Substation Event (GOOSE) messages. By integrating these physical and cyber elements, EPIC provides a realistic environment for analysing operational behaviour, conducting cyber-attack simulations, and evaluating defence strategies in smart grid systems. Building on these capabilities, an attack scenario targeting the parallel operation of generators in EPIC is reported in~\cite{roomi2023analysis}.

\subsection{Sub-transmission level substation model:}

\begin{figure}[!h]
\centering
{\includegraphics[width=\textwidth]{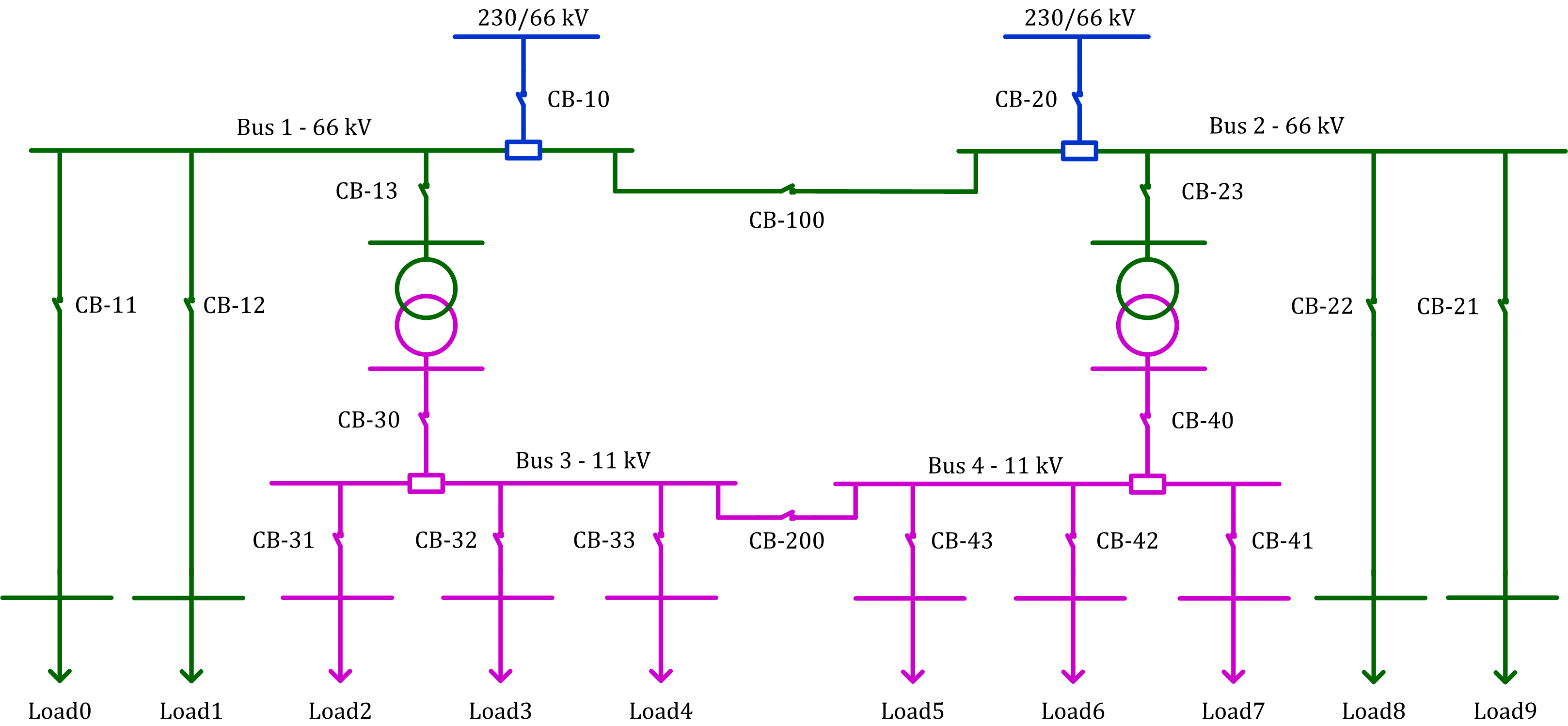}}
\caption{Single Line Diagram of 66/11 kV Substation Model}
\label{adsc}
\end{figure}

The substation modelled in this work operates at a sub-transmission voltage level of 66 kV, which is stepped down to an 11 kV distribution voltage level for downstream delivery. The electrical quantities are acquired through a range of field devices, including current transformers (CTs) for current measurement, voltage transformers (VTs) for voltage scaling, and transducers that convert analogue signals into digital forms suitable for monitoring and control. Switching states and protection actions are provided by circuit breakers (CBs), whose operational status is continuously communicated to the automation layer. These raw measurements and device states are transmitted to the programmable logic controllers (PLCs), human–machine interfaces (HMIs), and station servers via communication end points such as meters and intelligent sensors deployed across the network. The overall substation configuration is depicted in the single-line diagram shown in Fig.~\ref{adsc}. In addition to representing the hierarchical arrangement of primary equipment,~\cite{roomi2020false} details the inclusion of power, frequency, and temperature sensors, the connection and operational states of circuit breakers, and the integration of protection relays responsible for system safety. This comprehensive schematic enables a clear understanding of the measurement, protection, and control architecture within the substation, serving as the foundation for the experimental and analytical work presented in this study.

\subsection{3-substation system model:}

\begin{figure}[t]
\centering
{\includegraphics[width=\textwidth]{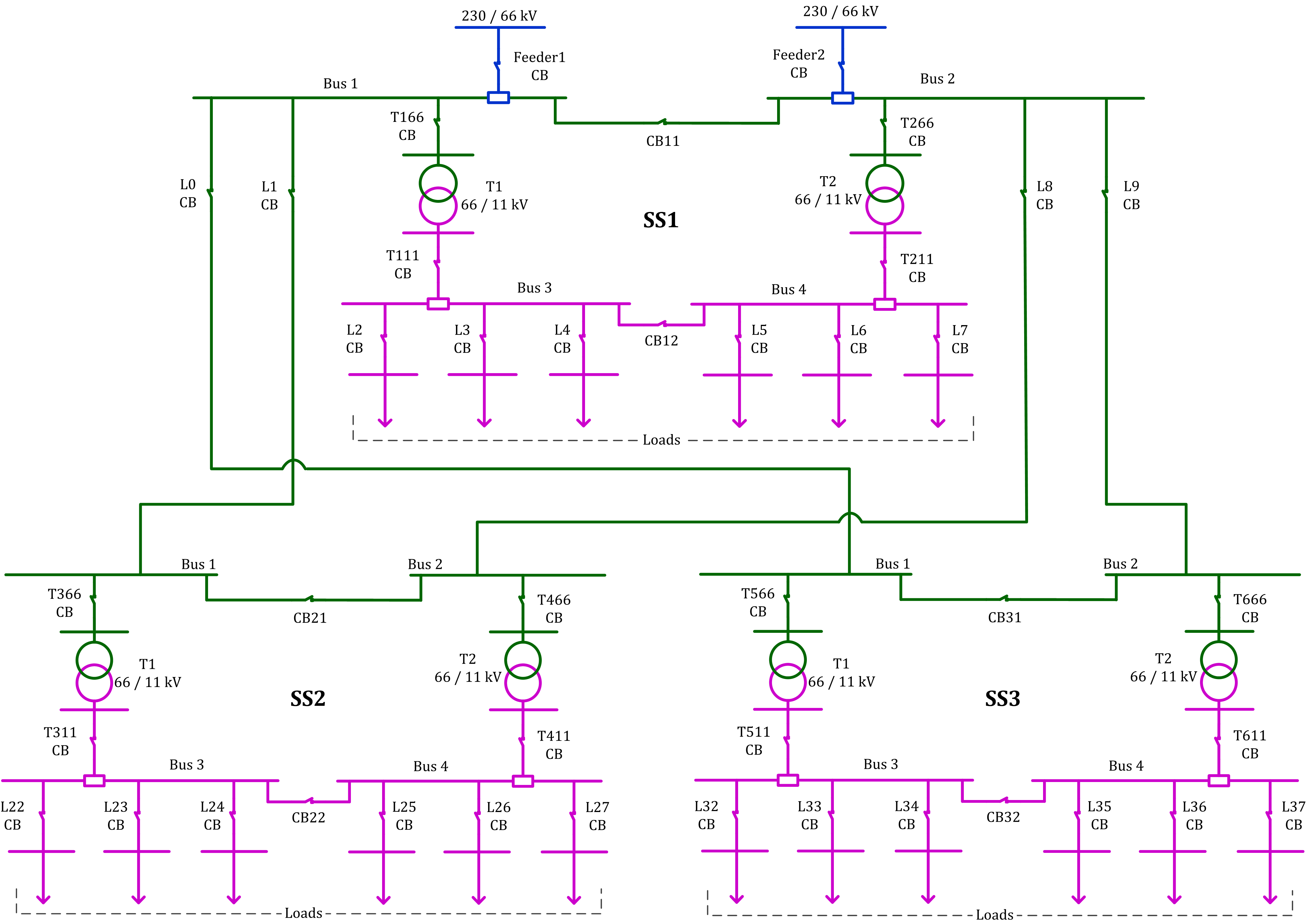}}
\caption{Single Line Diagram of 66/11 kV 3-Substation Model}
\label{adscmss}
\end{figure}

The three-substation model is developed as an extension of the single-substation configuration described in the previous subsection, enabling the study of larger-scale power system interactions. This model comprises three interconnected 66/11 kV sub-transmission substations (SS1, SS2, and SS3), forming a representative multi-substation network. The primary substation, SS1, is supplied by two incoming 66 kV feeders, which provide the main source of power to the system. From SS1, outgoing 66 kV transmission lines are extended and configured as incoming feeders to the secondary substations, SS2 and SS3, thereby establishing inter-substation connectivity. To enhance system resilience, redundant transmission lines are incorporated within the model to ensure continuity of supply and operational flexibility in the event of faults or emergency conditions. The overall electrical configuration and interconnection scheme are depicted in the single-line diagram of the three-substation model, as shown in Fig.~\ref{adscmss}, which illustrates both the hierarchical power flow and the redundancy mechanisms implemented in the network.

\section{Substation Configuration Files} \label{sec:scl}

\begin{center}
 \begin{tabular}{||p{3cm}||p{11cm}||} 
 \hline\hline
 Aim & Creating Substation Configuration Language Files\\ 
 \hline\hline
 Software & SCL Matrix~\cite{sclmatrix} \\ 
 \hline\hline
 Outcome & System Specification Description (SSD), IED Capability Description (ICD), System (Substation) Configuration Description (SCD) \\
 \hline\hline
\end{tabular}
\end{center}
SCL is the descriptive language defined by IEC 61850 for configuring electrical substations devices. The main parts of an SCL include Header, Substation, Communication, IED, Datatype templates.

\subsection{System Specification Description (SSD) file:}
\begin{figure}[h]
\centering
{\includegraphics[width=\textwidth, height=8cm]{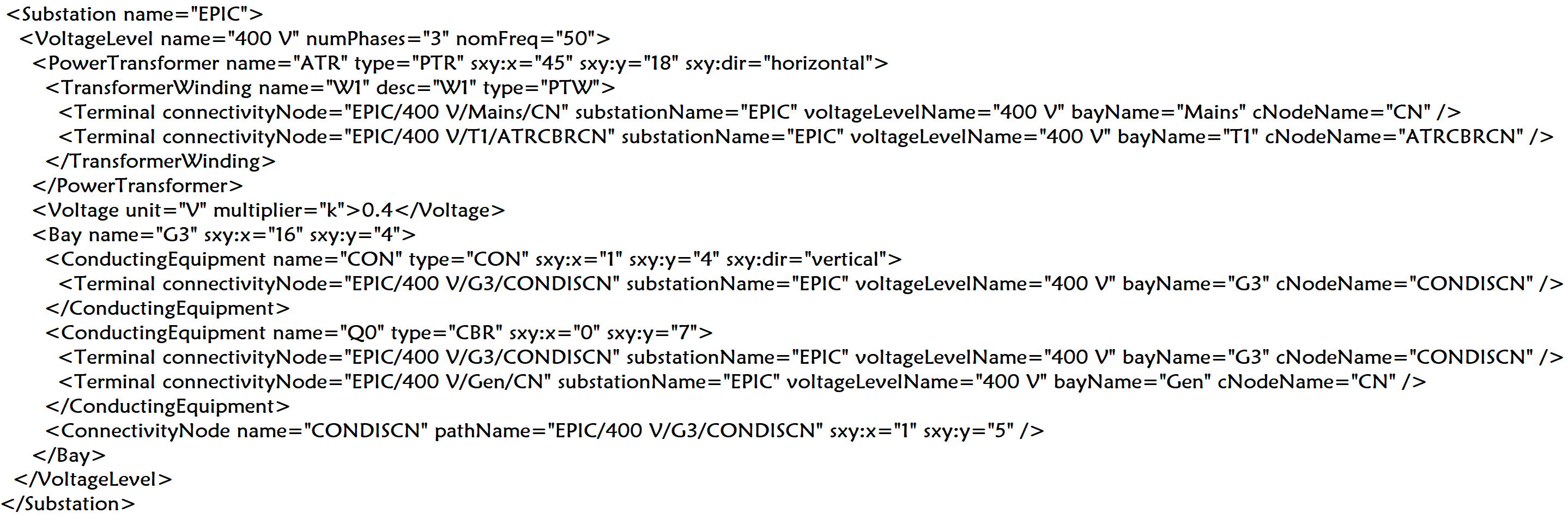}}
\caption {Description of SLD in SCL format.}
\label{ssd}
\end{figure}

This file includes the complete specification of the substation automation system (SAS). The single line diagram (SLD) of the SAS and its functionalities (allocation of logical nodes (LN) to the physical system components) are defined. The main component of the file is the ‘Substation’ section. A sample of the s
`Substation' section in an SSD file is shown in Fig.~\ref{ssd}.

The ‘Substation’ section in the SSD file includes the sub-section ‘Voltage Level’, which in turn consists of ‘Bay’. A single substation may contain one or more voltage levels and a single voltage level may contain one or more bays. Voltage level includes details such as operating voltage, number of phases, operating frequency, alternating transformer, etc. A Bay represents a single feeder line in the substation. Therefore, it includes the details of the physical components (represented as ‘ConductingEquipment’ in SCL) such as circuit breaker, relay, disconnectors, etc., that are present in the feeder line. A node is the point of connection between two physical components. The ‘Terminal connectivityNode’ in the above figure represents the location of a particular in the substation, with which the location of any physical components can be determined. The above SSD figure demonstrates the simple SCL structure of the EPIC testbed model without the allocation of Logical Nodes (LN) to the power system components. The LN can be assigned to the physical equipment after the integration of Intelligent Electronic Device (IED) description file to the substation. 

\subsection{IED Capability Description (ICD) file:}

IEDs are widely deployed in power automation systems due to their interoperability and integration flexibility. The primary function of an IED is to provide protection function. Some common types of IED include protective relay devices, circuit breaker controllers, voltage regulators, re-closer controllers, capacitor bank switches, tap change controllers, etc. IEDs receive measurements from sensor and other power equipment. According to the measurements, IEDs can either provide control commands to trip circuit breakers if any anomalies in voltage, current or frequency is detected or adjust the tap controllers to maintain the voltage at the desired level. The file which describes the functionalities of an IED is called the IED Capability Description (ICD) file, usually provided by the manufacturer based on the requirements. 

\begin{figure}[h]
\centering
 {\includegraphics[width=\textwidth, height=11cm]{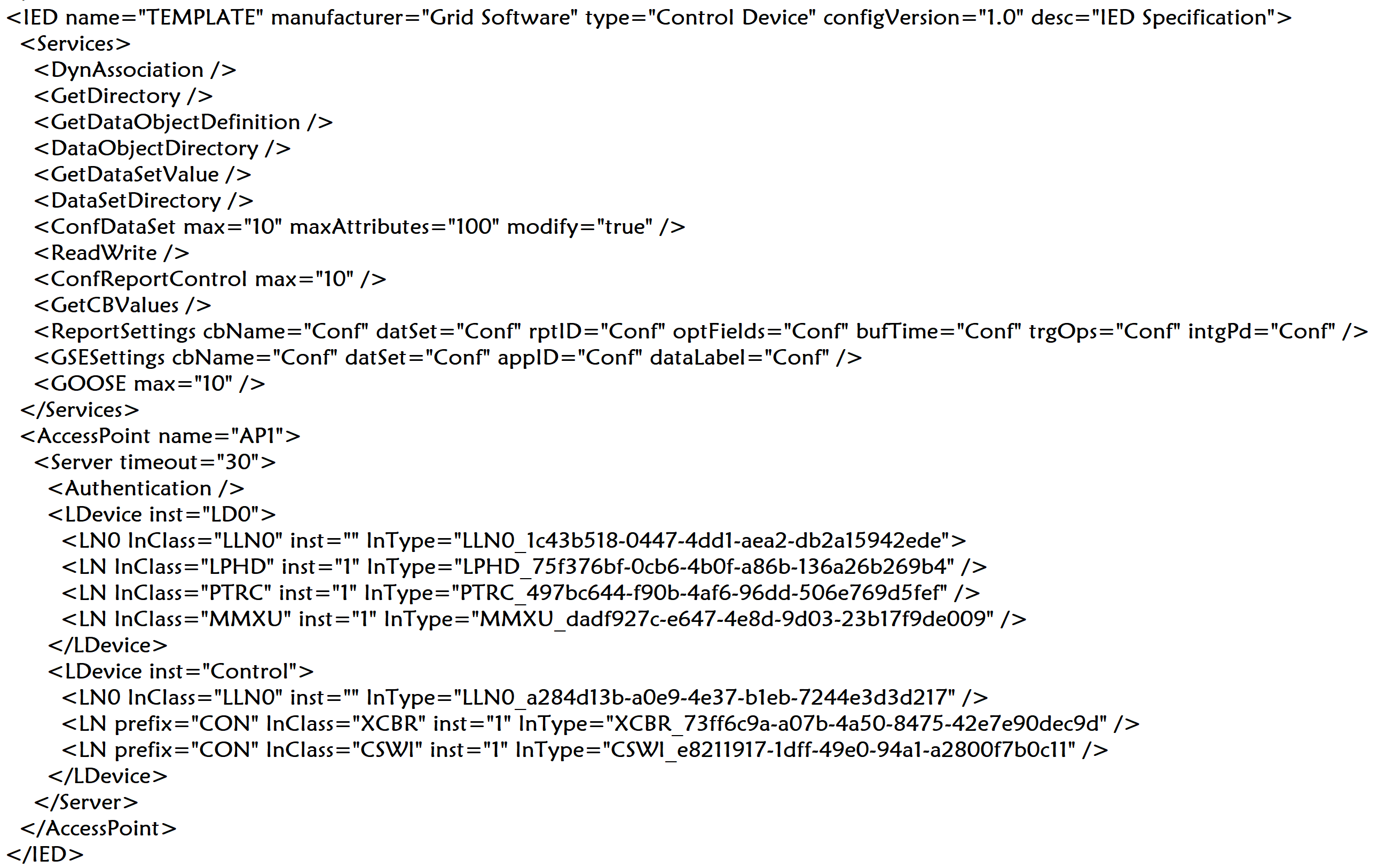}}
\caption{Description of ICD (IED section) in SCL format.}
\label{icd}
\end{figure}

The main section of the ICD file is the ‘IED’ section. The template for an IED is shown in Fig.~\ref{icd}. This section contains the services included for a particular IED. As depicted in the figure, IED section provides the details of the Logical Devices (LDs) and the Logical Nodes (LNs). Every LD may contain group of LNs. 

LNs in IEC 61850 is used to define the basic functions in an IED and it contains a group of Data Objects (DOs). IEC 61850 part 7-4 defines logical node type or class. All the logical node classes are grouped based on their functions. Among the various LN classes, the protection class and the measurement class are defined in the virtual IEDs that are utilised in this project. The protection functions include Over-current protection (PTOC), Over-voltage/Under-voltage protection (PTOV/PTUV), Under-frequency protection (PTUF), Reverse power flow protection (PDOP), Thermal over-load (PTTR). Similarly, the measurements of three-phase voltages, three-phase currents, active power and frequency contribute towards the measurement class, which is represented as ‘MMXU’. Additionally, a single IED may contain multiple instances any LN type/class and this is represented by an instance number added as a suffix (e.g. MMXU1, MMXU2).

\subsubsection{Control Blocks}

The communication between IEDs and Supervisory Control And Data Acquisition (SCADA) mostly follows Client-Server communication service. Manufacturing Message Specification (MMS) protocol is used in IEC 61850 to implement the communication between IEDs and SCADA. IED acts as the server, waiting for request from SCADA. The Client SCADA initiates the communication and requests to read data or control command, upon which IED responds to the corresponding request.

\begin{figure}[h]
\centering
 {\includegraphics[width=\textwidth, height=10cm]{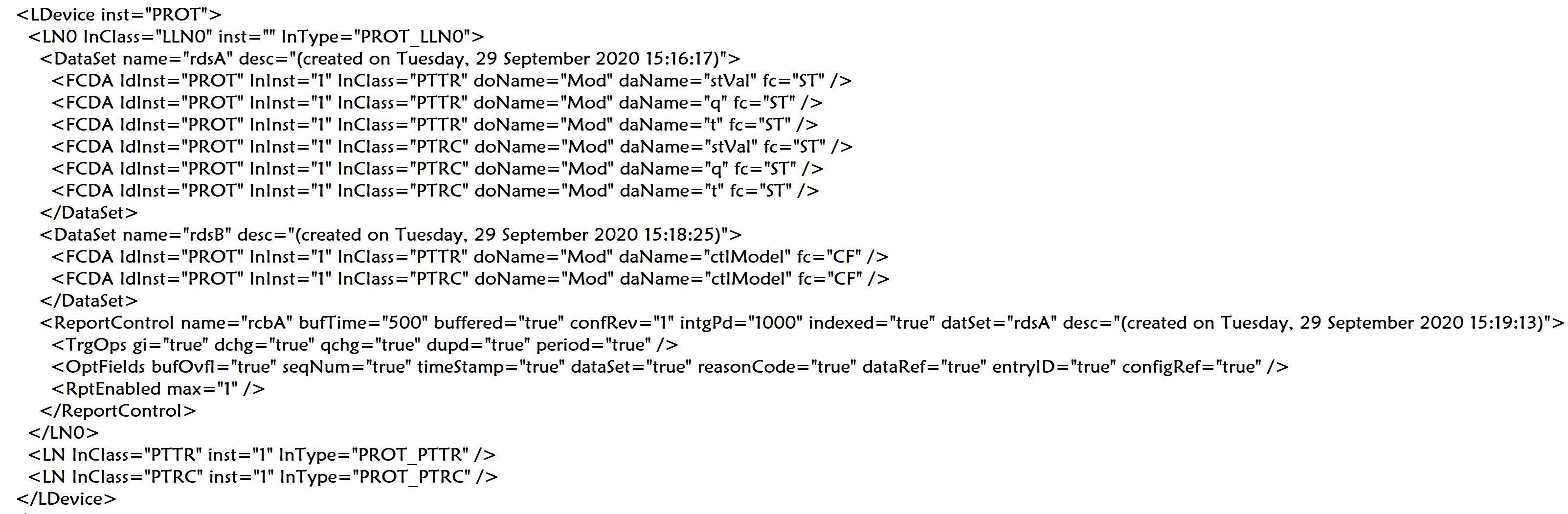}}
\caption{Description of ICD (Report Control Blocks) in SCL format.}
\label{icdrcb}
\end{figure}

In order to log the report event operation, IEDs need to be configured to prepare datasets containing the necessary data points and the report control blocks to specify how the data has to be communicated. Fig.~\ref{icdrcb} demonstrates the details of the report control block that includes the data related to the protection functionalities (thermal over-load and protection trip conditioning). Additionally, data objects related to measurements can be included in the control block. 

Generally, the communication to send event reports between IEDs and SCADA are established when there is a change in the data or quality attribute. However, periodical communication even when there is no change in the attributes can also be configured.

\subsubsection{GOOSE Control Block}

Generic Object Oriented Substation Events (GOOSE) communication is used in IEC 61850 to establish communications between IEDs. This communication follows the Publish - Subscribe service. The GOOSE-sending IED publishes the data to all IEDs in the network. However, the GOOSE data being published is retrieved by the IED that has subscribed and utilises the data.

\begin{figure}[b]
\centering
 {\includegraphics[width=\textwidth, height=5cm]{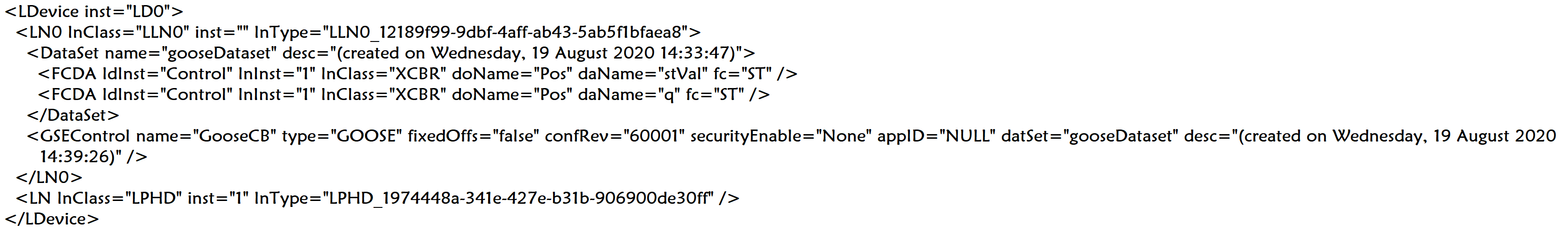}}
\caption{Description of ICD (GOOSE Control Blocks) in SCL format.}
\label{icdgcb}
\end{figure}

This communication involves at least one GOOSE-sending IED and one or more GOOSE-receiving IEDs. Similar to the report control blocks, data points that are to be published are created as a dataset. Subsequently, a GOOSE control block is configured that specifies how the GOOSE message has to be communicated. Fig.~\ref{icdgcb} illustrates the GOOSE control block with the GOOSE dataset. This dataset includes the data points related to the circuit breaker position.

The configuration of GOOSE communication can be described in three steps: 
\begin{itemize}
  \item Setting up GOOSE-sending IED (data points grouped in to datasets and configuring GOOSE control block)
  \item Setting up the GOOSE-receiving IEDs
  \item Configuring the GOOSE-receiving IED by mapping the GOOSE data to the user logic of the IED
\end{itemize}

\subsubsection{Communication Section}

\begin{figure}[h]
\centering
 {\includegraphics[width=\textwidth]{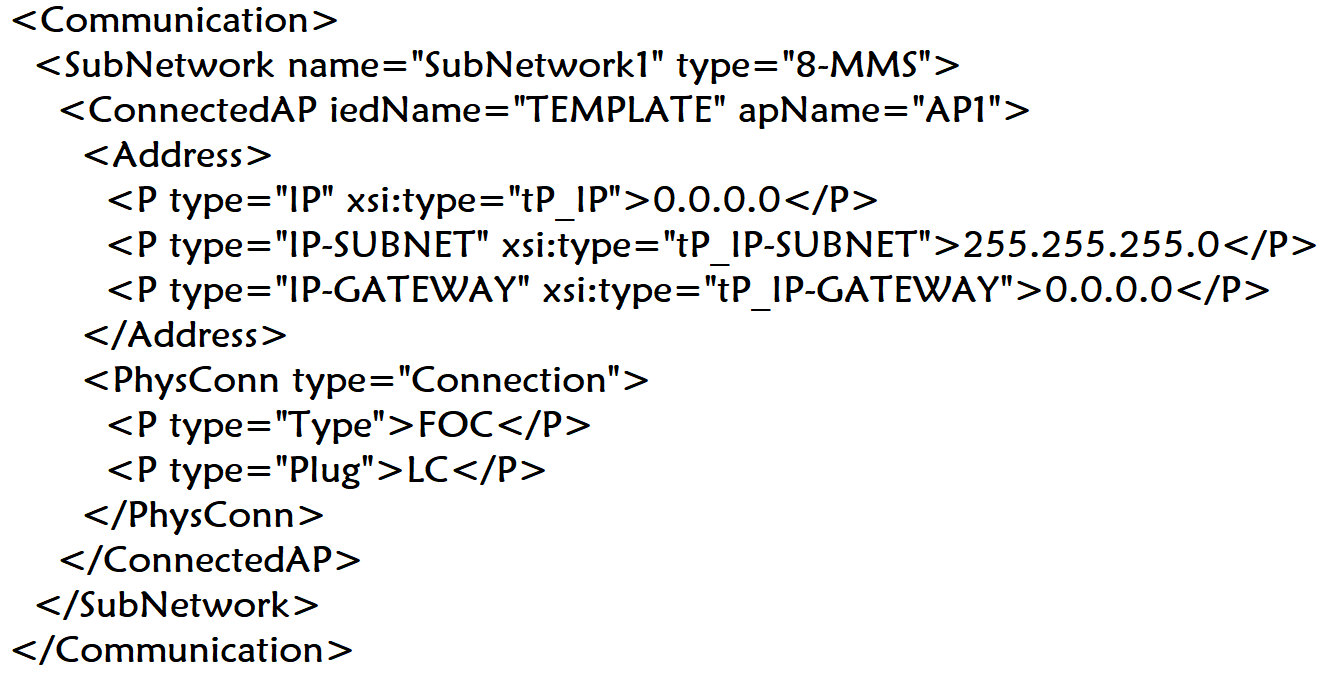}}
\caption{ICD (Communication Section) in SCL format.}
\label{icdcom}
\end{figure}

One another important section in the ICD file is the ‘Communication Section’. This section contains the details of the communication services supported by the IED. Fig.~\ref{icdcom} represents the communication section in the ICD file. It incldues details such as the SubNetwork name, communication protocol, IP address details, Physical Connection features.

\subsection{System (Substation) Configuration Description (SCD) file:}

The SSD and ICD files are processed by the System Configurator Tool (SCT), the resultant of which is the System (Substation) Configuration Description (SCD) file. This file includes the physical and cyber aspects of the entire substation. Once the processing is done, the resultant SCD files will have additional details, when compared with the individual SSD and ICD files. The changes in the configuration are listed in the following subsections.

\subsubsection{Configured SSD in SCD file}

\begin{figure}[h]
\centering
 {\includegraphics[width=\textwidth]{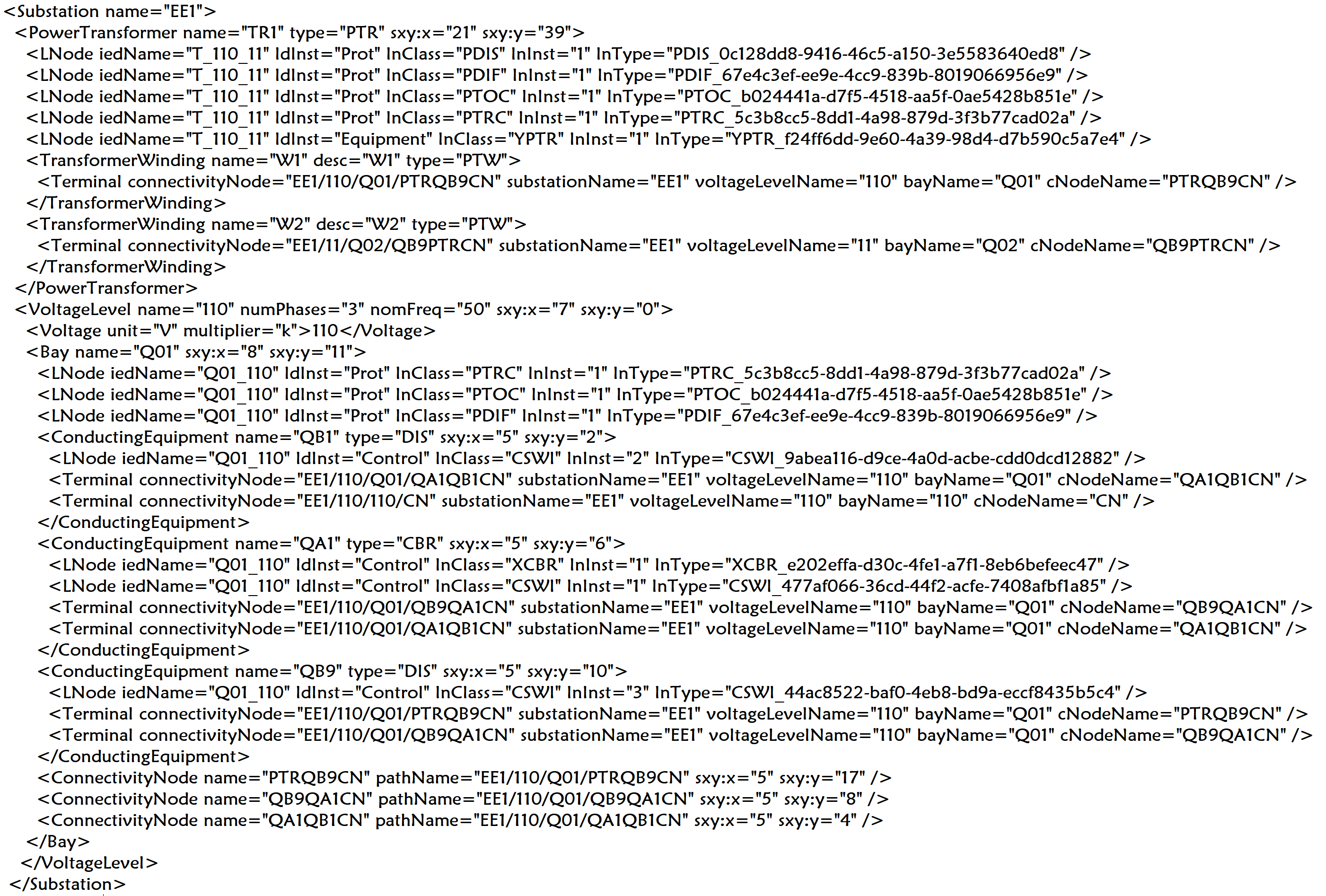}}
\caption{Configured SSD in SCL format.}
\label{ssdscd}
\end{figure}

Fig.~\ref{ssdscd} exemplifies the details included in the ‘Substation’ section of the SSD file. In the SCD file, the Logical Nodes in the ICD files are allocated to the physical components in the SSD file. Therefore, attributes associated with measurements, protection, switch gear, etc. are defined in the SCD files. For e.g. consider the power transformer (PowerTransformer name=“TR1”) in the figure. The choice of which specific LN is allocated to the power transformer from a particular IED is one among the details realised. Subsequently, the physical location of the physical component in the substation is also described.

\begin{figure}[!h]
\centering
 {\includegraphics[width=\textwidth, height=\textheight]{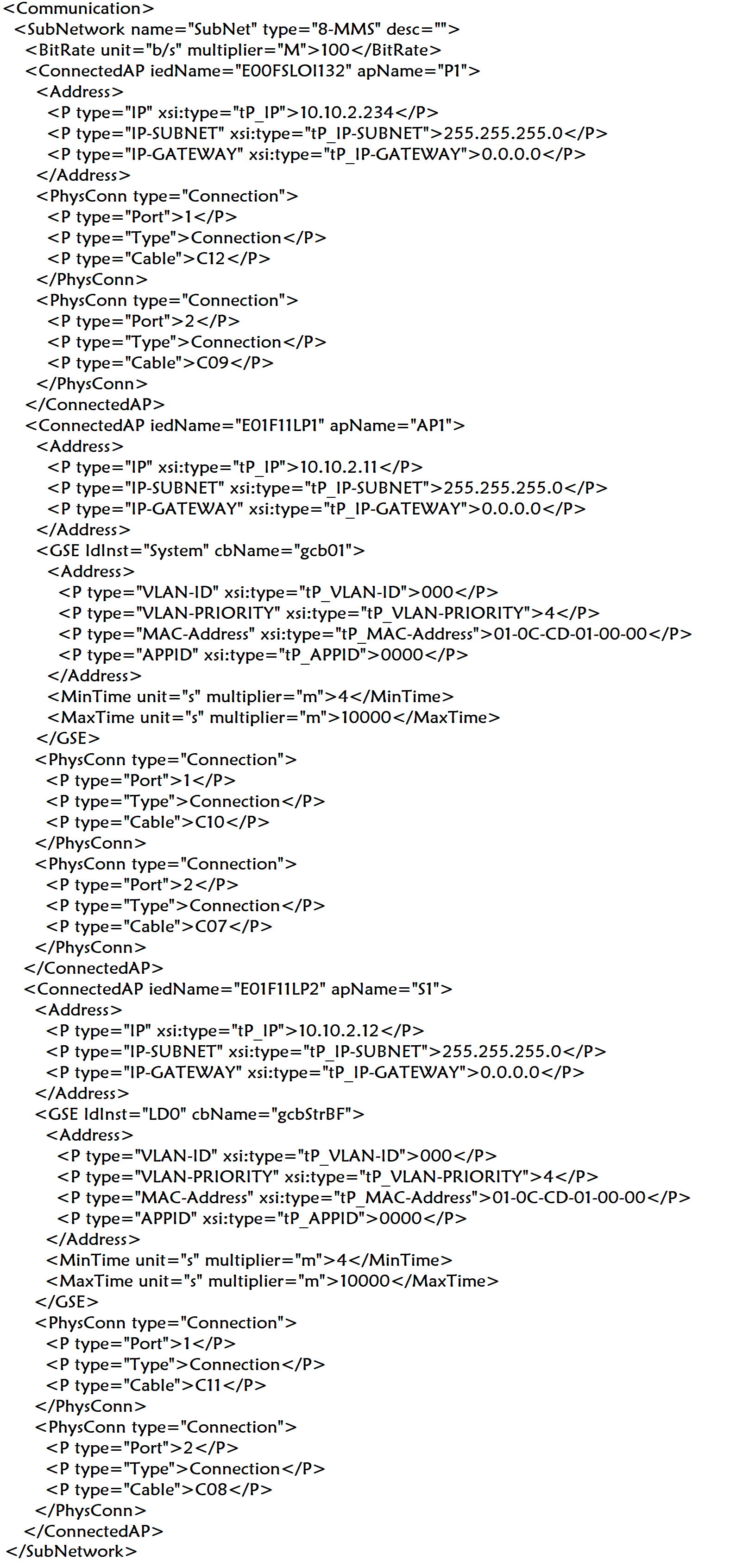}}
\caption{CID (Communication Section) in SCL format.}
\label{cidcom}
\end{figure}

\subsubsection{Configured IED description (CID) in SCD file}
Once the ICD files are utilised to build the SCD file using a SCT, details of the communication between the IEDs, and the communication between IEDs and SCADA systems are updated in the SCL files. The details of this can be extracted from the SCT as Configured IED description (CID) files.  Fig.~\ref{cidcom} demonstrates the communication section extracted from the SCD files. As seen in the figure, the SCL code initially contains the SubNetwork name and details. Subsequently, the connected access points (AP) of each IEDs in the substation system are displayed. Each ‘ConnectedAP’ section lists the details of the IP address, Generic Substation Event (GSE) details and Physical Connection details. GSE details include the DataPoints that are to be published by a particular IED in the form of GCB, and also the MAC-address details of the IED. The Physical Connection contains the details of the ports through which the IEDs are connected between each other and to the SCADA/HMI (Human Machine Interface). 

\subsection{System/Substation Exchange Description (SED) file:}

The SG-ML utilizes the IEC 61850 SED (System Exchange Description) file for configuring the IEDs in both the substations for enabling the inter-substation communication. Prior to configuring inter-substation communication, the individual substations are already configured and their SCD files are available.

Utilizing the SCD files of two substations SED file is created. The SED file contains the information related to IEDs involved in inter-substation communication from both the substations. The SED file contains information about the electrical connection between the two substations and the communication network information, control blocks and semantics (LN model) of IEDs involved in inter-substation communication.

Using the required portions from the SED file and a proprietary XML file a virtual cyber range for inter-substation communication is developed. The proprietary XML file contains the necessary additional parameters required for power system and cybernetwork simulation.

\begin{figure}[h]
\centering
 {\includegraphics[width=\textwidth]{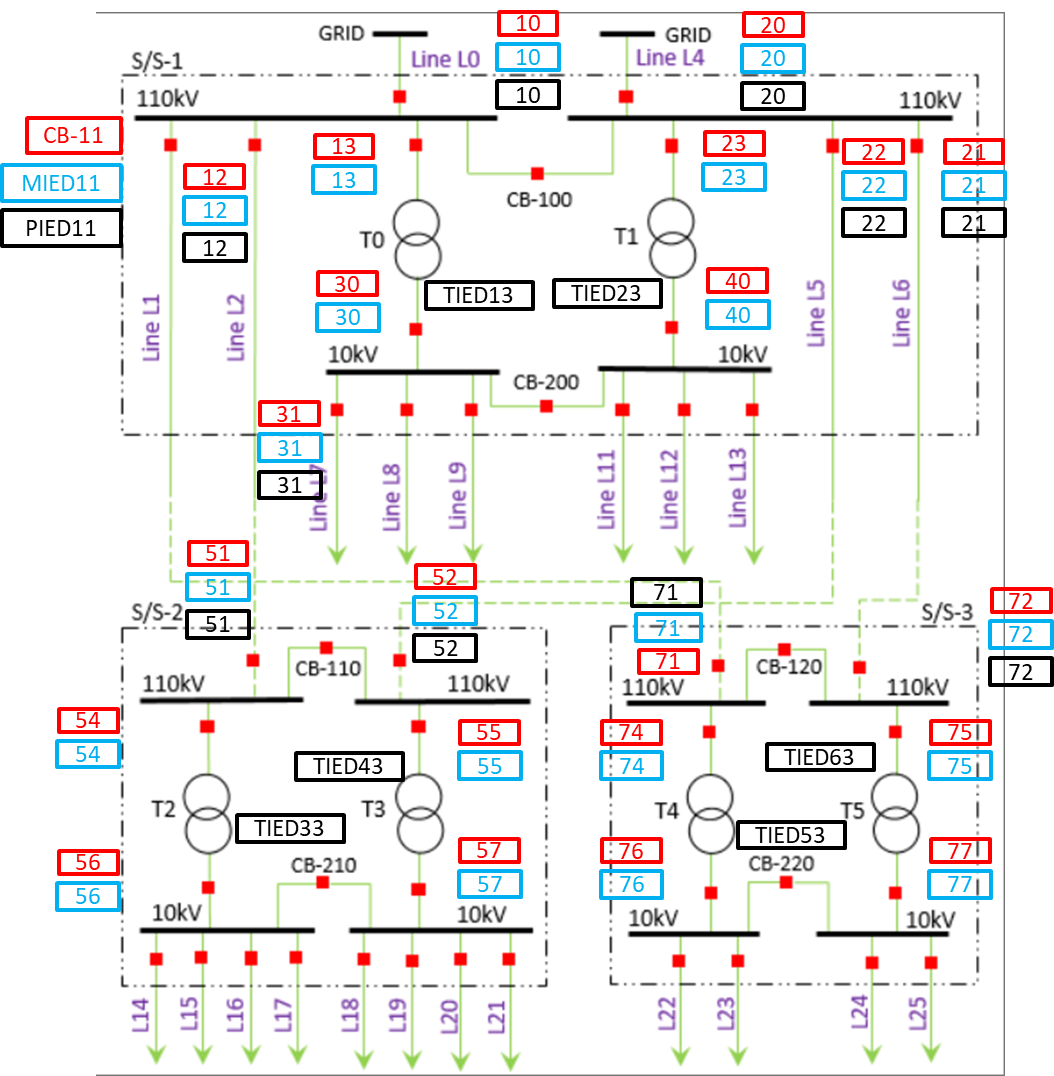}}
\caption{Topology of three substation model and the IEDs associated.}
\label{3substation}
\end{figure}

Example:

Fig.~\ref{3substation} shows the topology of three substation model and the IEDs associated to it. An SED file for inter-substation communication between substation 1 (S-1) and substation 2 (S-2) over Line L2 is considered and discussed here. Differential protection scheme is considered on Line L2. The MIED12 sends the sample measured values to PIED12 and PIED51. Similarly, the MIED51 sends the sample measured values to PIED51 and PIED12.

\begin{figure}[h]
\centering
 {\includegraphics[width=\textwidth]{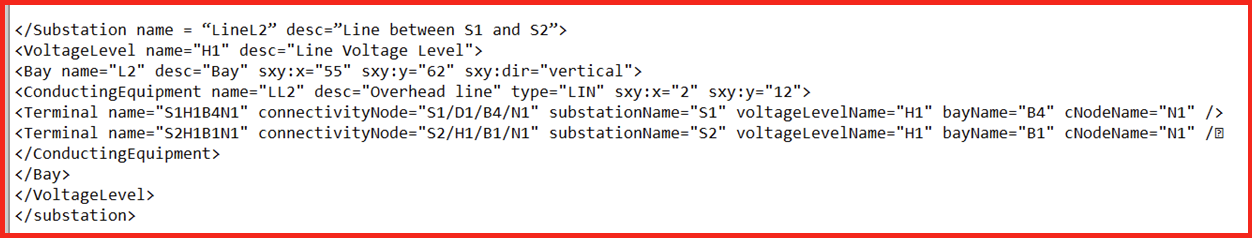}}
\caption{Portion of SED file related to the electrical connection and terminal nodes between two substations.}
\label{nodesS1S2}
\end{figure}

\begin{figure}[h]
\centering
 {\includegraphics[width=\textwidth]{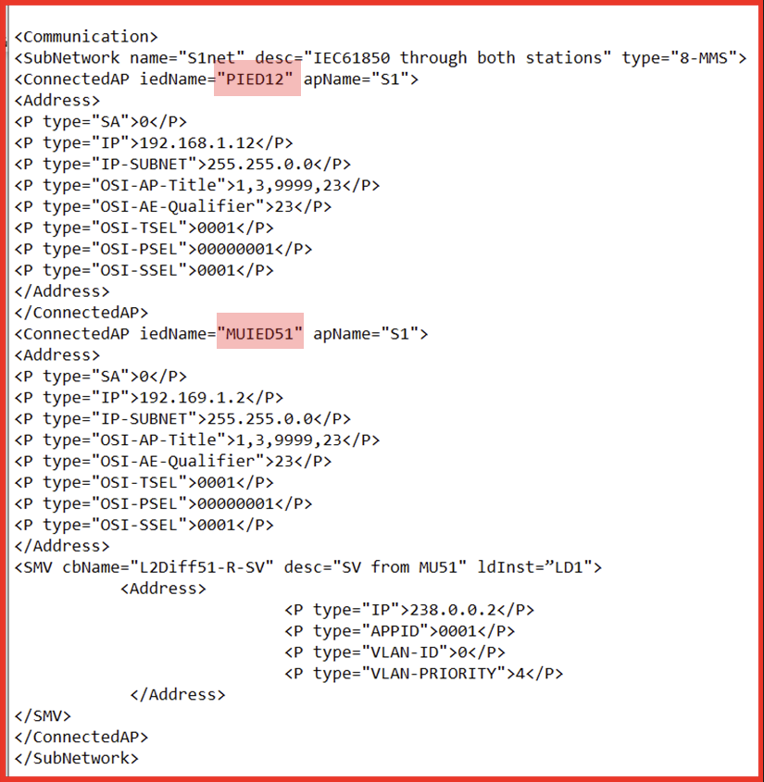}}
\caption{Portion of SED file related to Subnet communication information of substation 1 with info of IEDs from substation 2.}
\label{commS1S2}
\end{figure}

\begin{figure}
\centering
 {\includegraphics[width=\textwidth, height=\textheight]{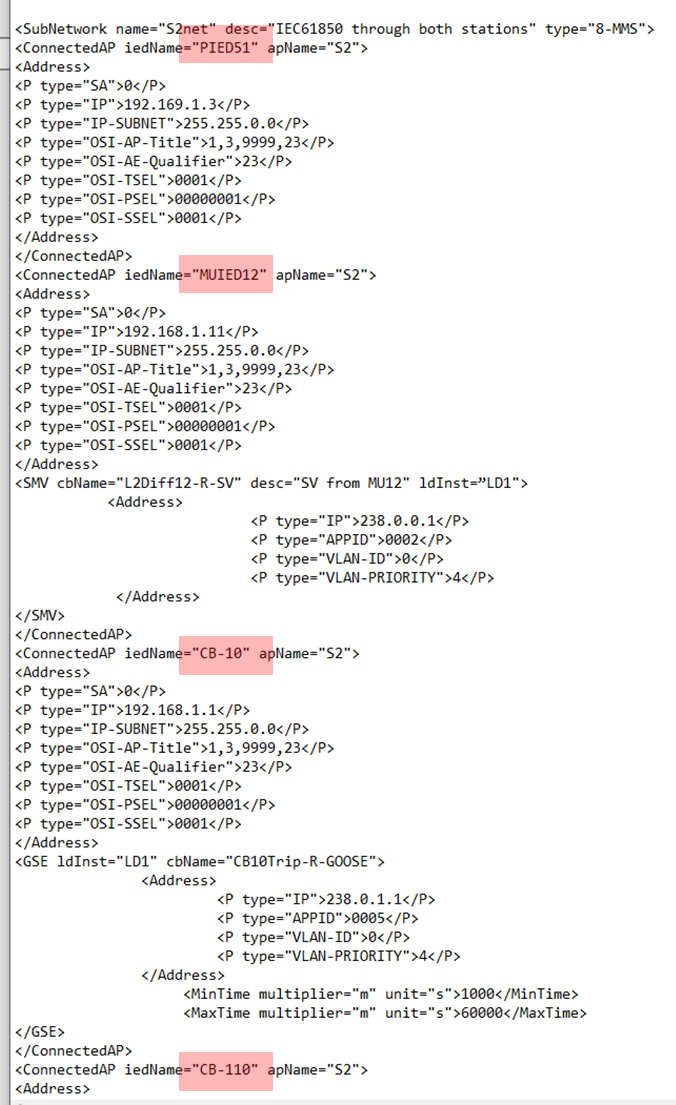}}
\caption{Portion of SED file related to Subnet communication information of substation 1 with info of IEDs from substation 2.}
\label{IEDS1S2}
\end{figure}

The main parts of the SED file for the inter-substation communication is shown in Figs.~\ref{nodesS1S2},~\ref{commS1S2} and~\ref{IEDS1S2}. Fig.~\ref{nodesS1S2} shows the part of SED file related to the electrical connection and terminal nodes between two substations. 
Figs.~\ref{commS1S2} and~\ref{IEDS1S2} shows the communication portion of SED file. Fig.~\ref{commS1S2} shows the Subnet of substation 1 which contains additional information of the IEDs from substation 2 such as the network information, control blocks and semantics (LN model) of IEDs from substation 2. Similarly, the Fig.~\ref{IEDS1S2} shows the Subnet information of substation 2 containing information of IEDs from substation 1.

\section{SCL (SSD) to XML Schema} \label{sec:sgml1}
\begin{center}
 \begin{tabular}{||p{4cm}||p{10cm}||} 
 \hline\hline
 Aim & System Specification Description to eXtensible Markup Language\\ 
 \hline\hline
 Programming Language & Python~\cite{python} \\ 
 \hline\hline
 Outcome & SCL Processor to generate Power System Simulation Model \\
 \hline\hline
\end{tabular}
\end{center}

\begin{figure}[h]
 \centering{\includegraphics[width=.7\textwidth, height=3cm]{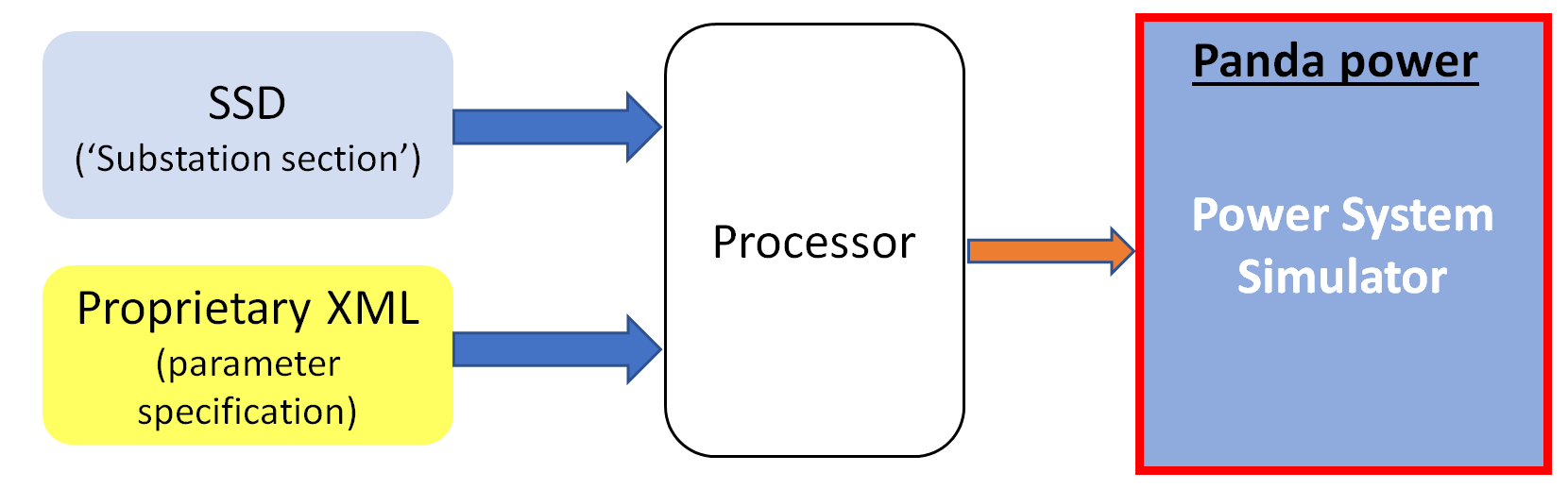}}
\caption{Framework for SSD to XML process.}
\label{ssdxmlfc}
\end{figure}

In the section, the focus is on creating an XML schema from the SSD file using parser in Python. The framework for the process is depicted in Fig.~\ref{ssdxmlfc}.

In order to achieve this two files are required. As depicted in Fig.~\ref{fig:sgml}, one file is an SCL file and another is an proprietary XML file. The ‘Substation’ section in the SSD file describes the physical components and their connections. However, this file does not include the parameter specification of the physical components in the substation. Therefore, an additional proprietary XML file is created that describes the specification of each component in the substation. Merging these two files via a parser in Python, the outcome would be a processor that process the SCL and the proprietary file to generate the power system simulation model. The following figures (Fig.~\ref{parssd} and Fig.~\ref{parpropxml}) show the LN related to a physical component in SSD and their parameter specification in the corresponding proprietary XML files, respectively. 

\begin{figure}[h]
\centering
 {\includegraphics[width=\textwidth, height=10cm]{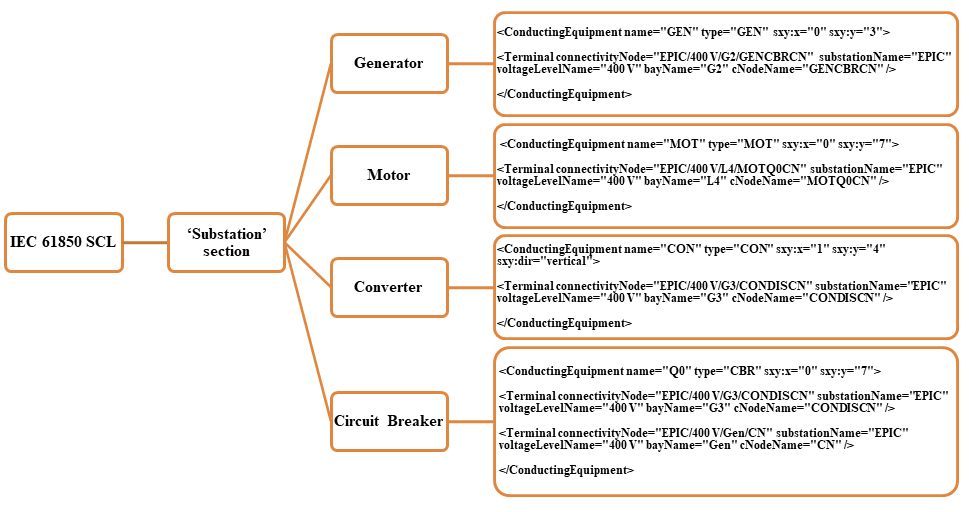}}
\caption{IEC 61850 SCL (‘Substation’ section).}
\label{parssd}
\end{figure}

\begin{figure}[h]
\centering
 {\includegraphics[width=\textwidth, height=9cm]{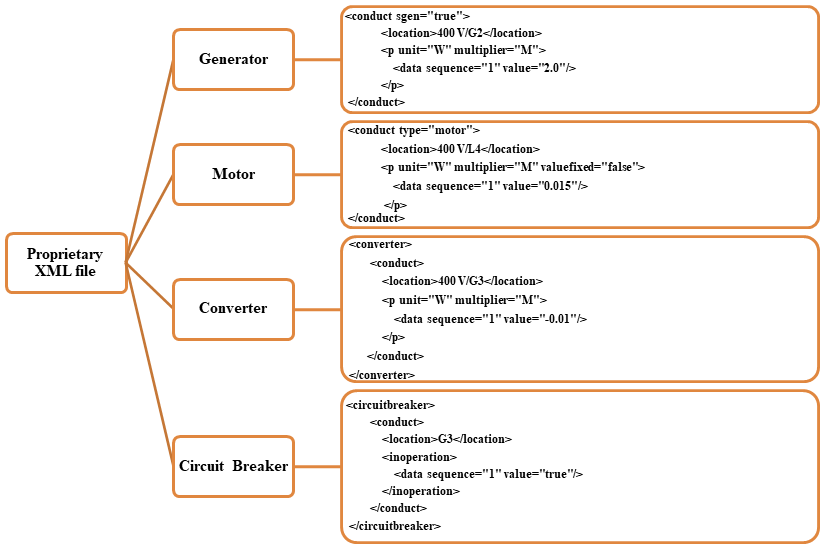}}
\caption{Proprietary XML (parameter specification).}
\label{parpropxml}
\end{figure}

\section{Virtual IED} \label{sec:sgml2}
\begin{center}
 \begin{tabular}{||p{4cm}||p{10cm}||} 
 \hline\hline
 Aim & XML Schema for ICD and Threshold Logic\\ 
 \hline\hline
 Software & Eclipse~\cite{eclipse}  \\ 
 \hline\hline
 Outcome & Virtual IED \\
 \hline\hline
\end{tabular}
\end{center}

The following requirements should be addressed to create a virtual IED.
\begin{itemize}
    \item Creating an XML schema for the input ICD files
    \item Creating an XML schema for the Proprietary settings file
\end{itemize}

\begin{figure}[h]
 \centering{\includegraphics[width=.8\textwidth, height=5cm]{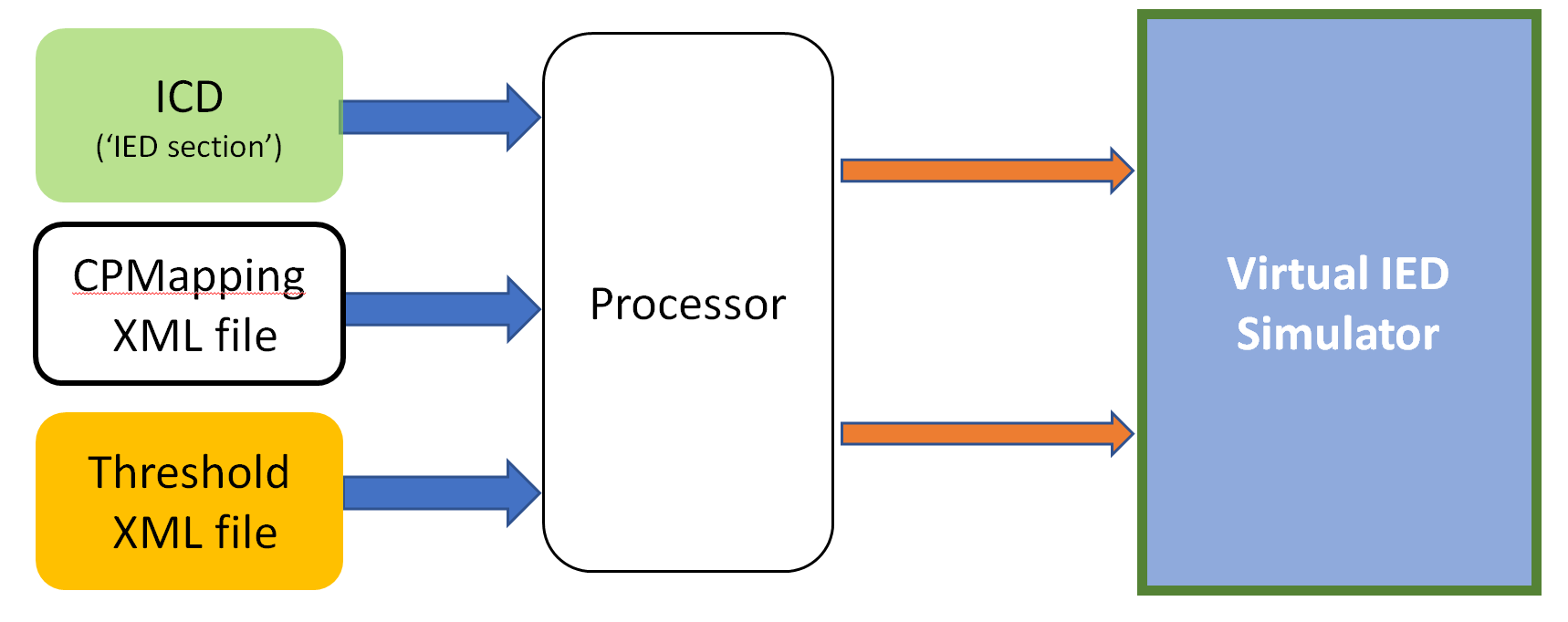}}
\caption{Framework for Virtual IED process.}
\label{iedxmlfc}
\end{figure}

The framework for the aforementioned process is depicted in Fig.~\ref{iedxmlfc}.

\subsection{XML schema for input ICD files}

\begin{figure}[hb!]
\centering
 {\includegraphics[width=\textwidth]{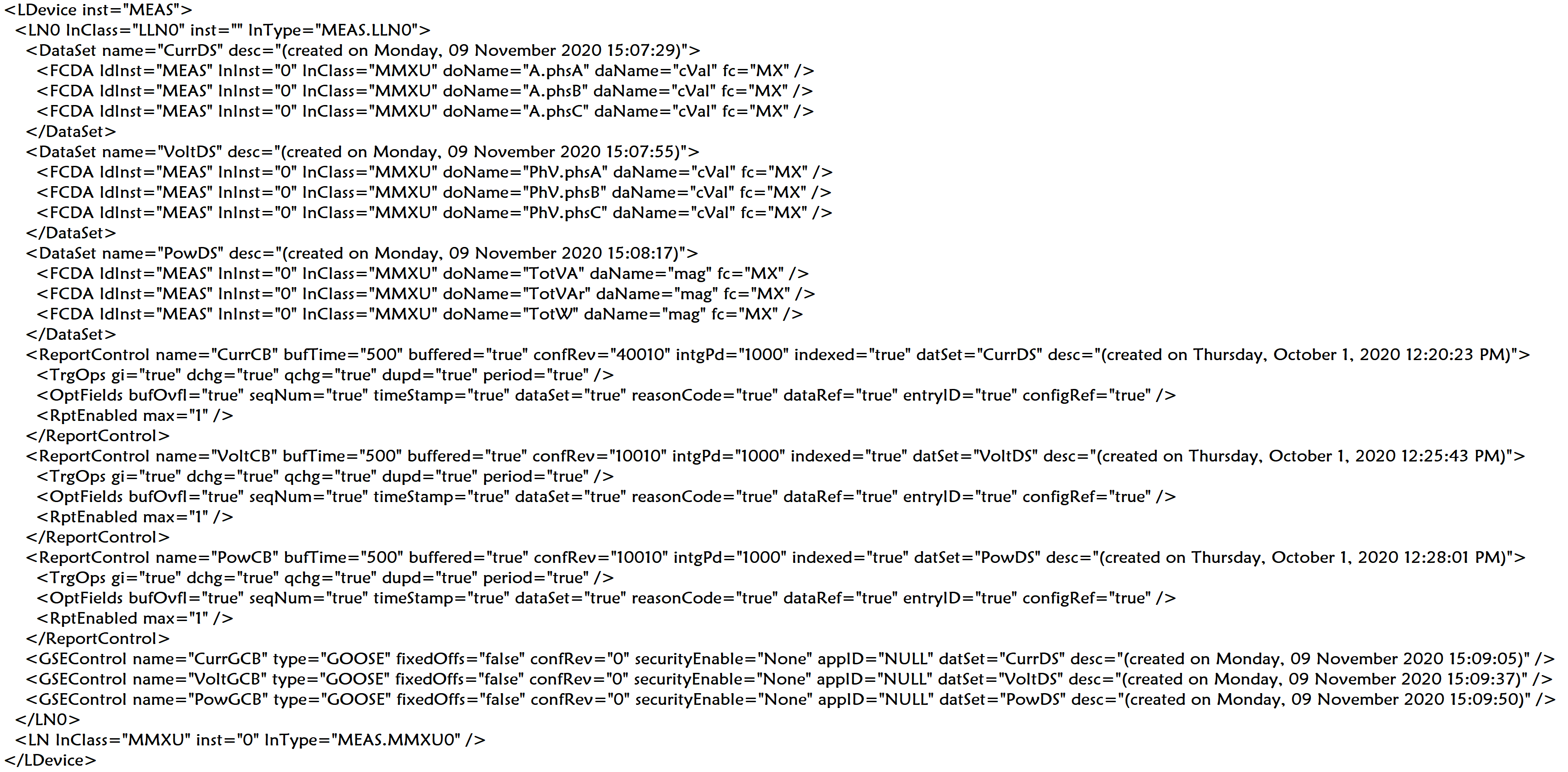}}
\caption{Input ICD file for creating XML schema.}
\label{icdin}
\end{figure}

In order to create the XML schema for input ICD files, the attributes that are needed for the project are initially identified. As such, an ICD file that contains the attributes related to physical measurements such as voltages, currents and power is created. Fig.~\ref{icdin} depicts the LD (‘MEAS’). This section contains the details of the data attributes associated with every LN instances, the data points that are used to create the datasets, and the allocation of datasets to the report control blocks. 

\begin{figure}[b]
\centering
 {\includegraphics[width=\textwidth]{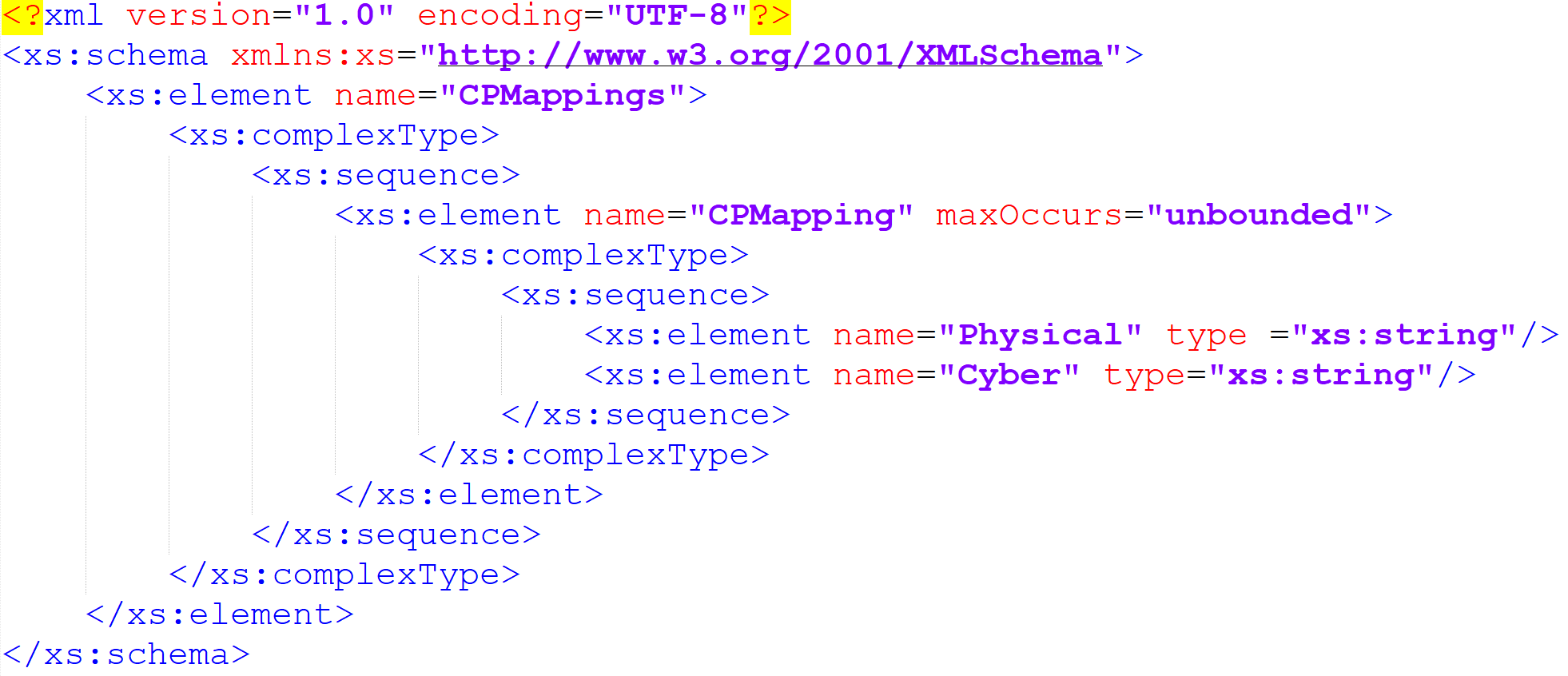}}
\caption{XML schema for cyber-physical mapping.}
\label{xmlout}
\end{figure}

\begin{figure}[!b]
\centering
 {\includegraphics{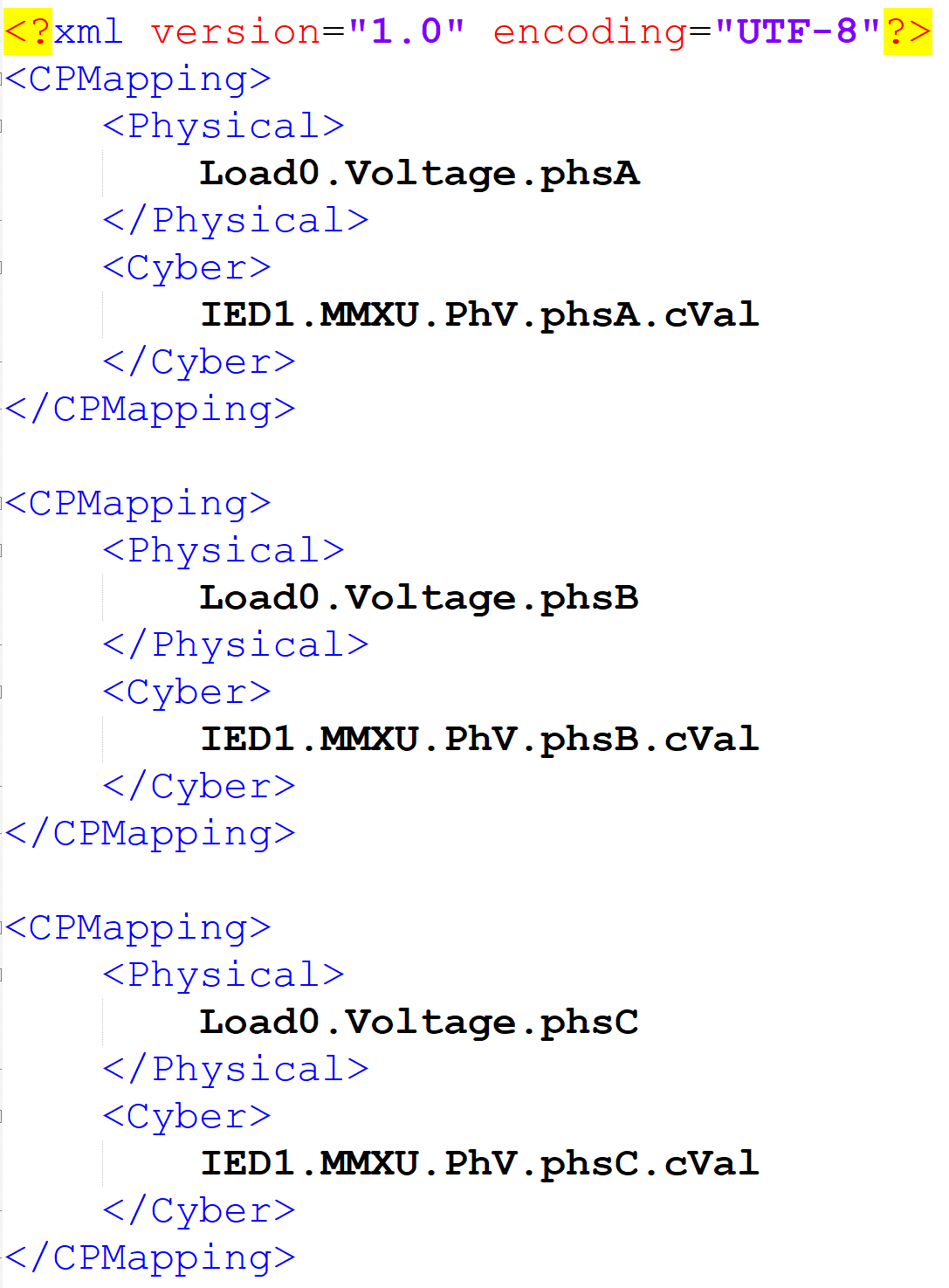}}
\caption{XML file for cyber-physical mapping.}
\label{xmlcp}
\end{figure}

From Fig.~\ref{icdin}, the attributes associated with each physical measurement can be derived. Subsequently, an XML schema has to be created, to map the physical measurements with the IED attributes. Fig.~\ref{xmlout} demonstrates the schema for mapping between the physical (real-time measurements) and cyber (IED attributes) aspects for a substation. An example of the XML file that describes the physical and cyber parameters for the system is shown in Fig.~\ref{xmlcp}. The example depicted in Fig.~\ref{xmlcp} represents the attributes related to ‘Load 0’ in Fig.~\ref{adsc}. The three-phase load voltages for ‘Load0’ are measured, and are assigned to physical attributes ‘Load0.Voltage.phsA’, ‘Load0.Voltage.phsB’ and ‘Load0.Voltage.phsC’. Now the three physical attributes are mapped to their cyber attributes ‘IED1.MMXU.PhV.phsA.cVal’, ‘IED1.MMXU.PhV.phsB.cVal’ and ‘IED1.MMXU.PhV.phsC.cVal’, respectively. These cyber attributes are present in the IED that has been connected to ‘Load0’ line in Fig.~\ref{adsc}.

\subsection{XML schema for Proprietary Settings file}

\begin{figure}[!h]
\centering
 {\includegraphics[width=\textwidth, height=21cm]{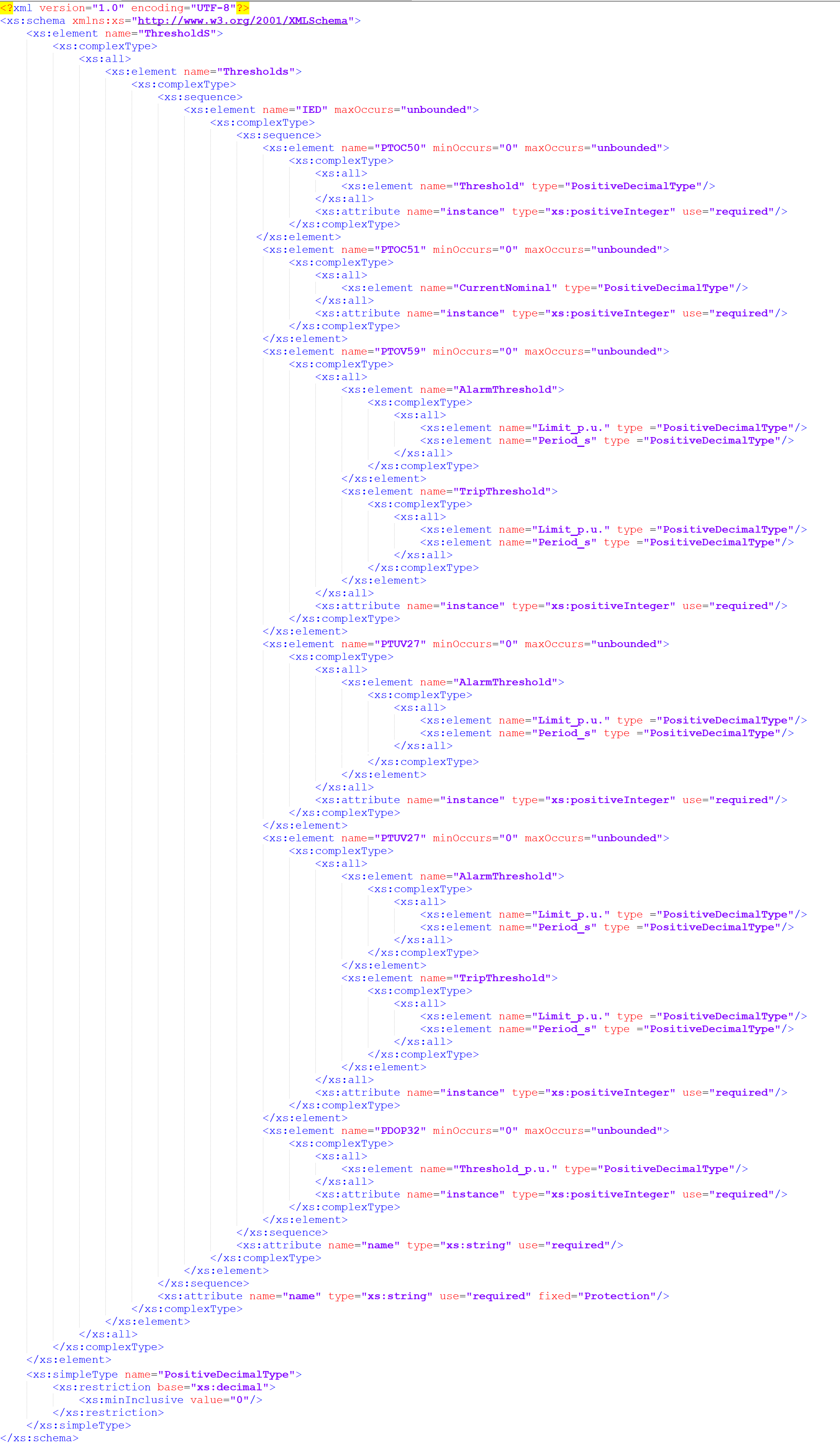}}
\caption{XML schema for Proprietary File (Thresholds Schema).}
\label{pfsch}
\end{figure}

\begin{figure}[!h]
\centering
 {\includegraphics{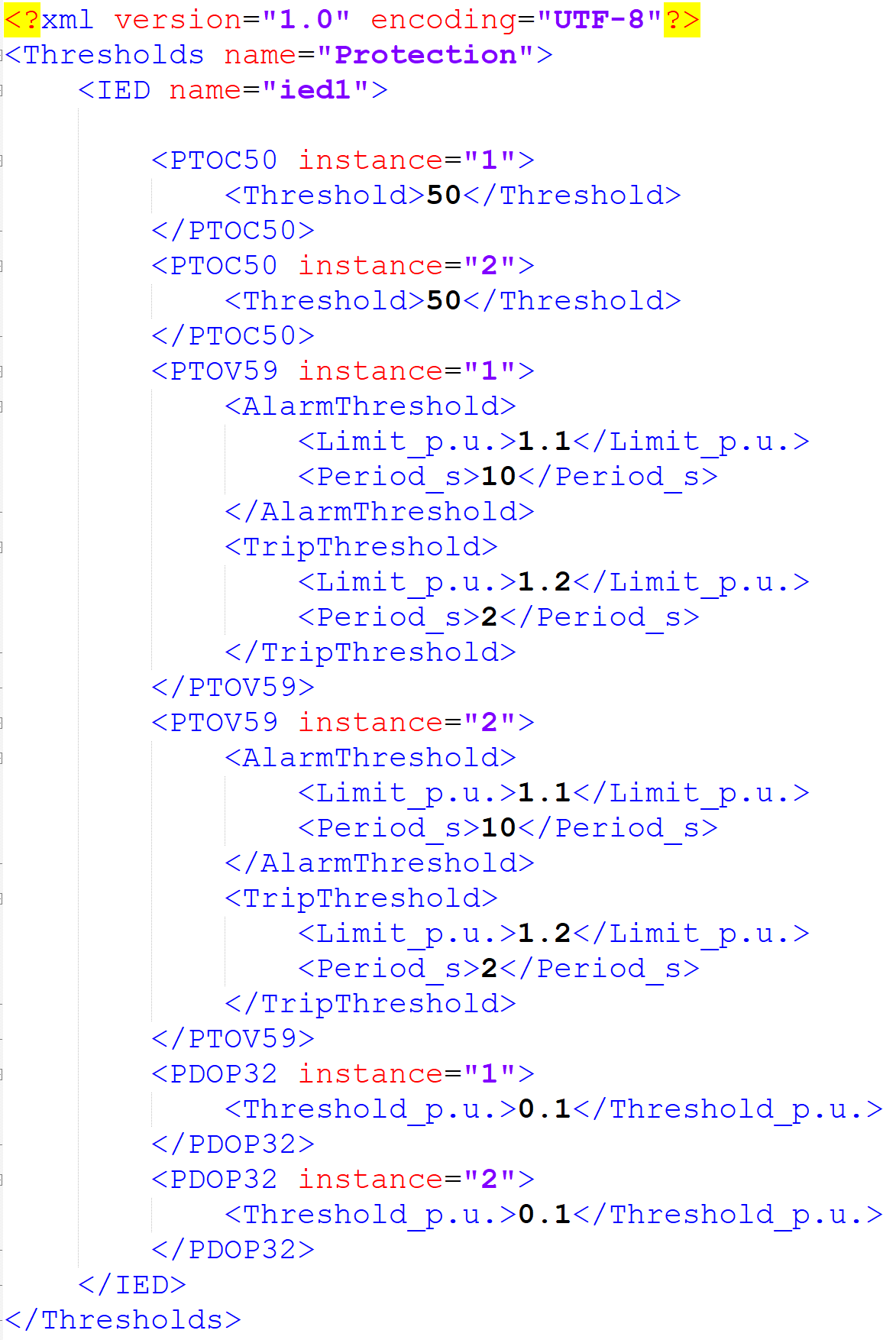}}
\caption{XML schema for Proprietary File (Thresholds Schema).}
\label{xmlpf}
\end{figure}

The implementation of virtual IED has to include the logic that is implemented for providing protection functionalities. Therefore, an XML schema that contains the threshold values has to be created. The logic considered in this schema are over over-/under-current, over-/under-voltage, reverse power protections. Fig.~\ref{pfsch} exemplifies the threshold conditions for each protection. As can be seen, two protection conditions have been implemented. The first condition is an ‘Alarm threshold’, the outcome would be an error message. The second condition is ‘Trip threshold’ at which point a trip command will be issued by the IED. Table.~\ref{psf} tabulates the threshold limits included in the proprietary settings file. An example of the XML file that describes the threshold settings for the proprietary settings file for the protections implemented in the IED is shown in Fig.~\ref{xmlpf}. Table.~\ref{txmlpf} tabulates the explanation of the attributes utilised in Fig.~\ref{xmlpf}.

\begin{table}[htbp]
\centering
\caption{Threshold limits for Proprietary Settings File}
    \begin{tabular}{|p{2cm}|p{4cm}|p{6.5cm}|}
    \hline
        Logical Nodes & Description & Threshold \\
        \hline
        PTOC (50) & Instantaneous over-current & 3 to 4 times * nominal current \\
        PTOC (51) & Time over-current & 1.05p.u. (starts picking up) \\
        PTOV (59) & Over-voltage & 1.1p.u. (10s) (alarm) \& 1.2p.u. (2s) (trip) \\
        PTUV (27) & Under-voltage & 0.8 p.u. (10s) (alarm) \& 0.7 p.u. (2s) (trip) \\
        PDOP (32) & Reverse Power & Any small amount (trip) \\
        \hline
    \end{tabular}
    \label{psf}
\end{table}

\begin{table}[H]
\centering
\caption{Attributes in XML Proprietary Settings file}
    \begin{tabular}{|p{4cm}|p{9cm}|}
    \hline
        Attributes & Description \\
        \hline
        ‘IED name’ & Represents the name of the IED \\
        ‘PTOC’, ‘PTOV’, ‘PDOP’ & IEC 61850 based protection \\
        ‘instance’ & Varies based on the number of devices in a load/line that requires a single type of protection\\
        ‘AlarmThreshold’ & Raises an alarm (message) in HMI \\
        ‘TripThreshold’ & Send ‘trip’ command to CB \\
        ‘p.u.’ & per-unit - the expression of system quantities as fractions of a defined base unit quantity\\
        ‘period’ & the time after which the IED sends a message or trip command\\
        \hline
    \end{tabular}
    \label{txmlpf}
\end{table}


\section{IEC 61131 PLCOpen} \label{sec:sgml3}
\begin{center}
 \begin{tabular}{||p{4cm}||p{10cm}||} 
 \hline\hline
 Aim & Virtual Programming Logic Controller\\ 
 \hline\hline
Standard \& Software & PLCOpen XML~\cite{iec61131} \& OpenPLC61850~\cite{roomi2022openplc61850} \\ 
 \hline\hline
 Outcome & Virtual PLC Auto-configurator \\
 \hline\hline
\end{tabular}
\end{center}

\begin{figure}[h]
\centering
 {\includegraphics{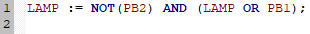}}
\caption{Structured Text representation in OpenPLC editor.}
\label{stedit}
\end{figure}

\begin{figure}[h]
\centering
 {\includegraphics[width=.8\textwidth, height=8cm]{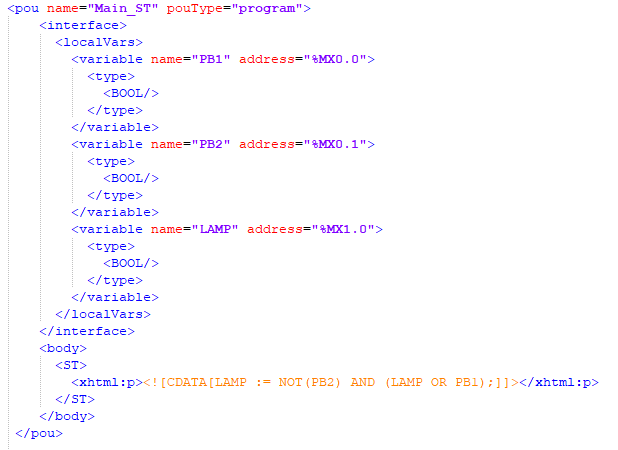}}
\caption{PLCopen XML: Structured Text.}
\label{plcopenxml}
\end{figure}

Building upon the work of~\cite{openplc}, the OpenPLC61850 project by~\cite{roomi2022openplc61850} enhances the existing OpenPLC software with support for the IEC 61850 protocol. At the core of this enhancement is SG-ML, which utilizes the industry-standard PLCOpen XML schema for representing PLC logic. This standardized XML format enables the flexible modeling of control logic and is independent of proprietary development tools. To illustrate, Fig.~\ref{stedit} shows the OpenPLC editor, from which the underlying PLC logic in structured text is extracted and represented in PLCOpen XML (Fig.~\ref{plcopenxml}). This process demonstrates the feasibility of using an open-source framework for advanced industrial control logic.

\begin{figure}
\centering
 {\includegraphics[width=.7\textwidth, height=4cm]{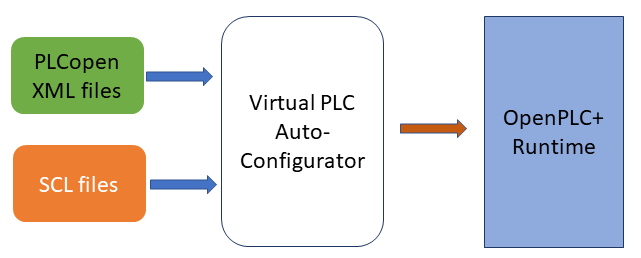}}
\caption{Virtual PLC Auto-configurator.}
\label{virtualplc}
\end{figure}

The ongoing research aims to extend the tool's capabilities by enabling the representation of PLCOpen XML from additional IEC 61131-3 languages, such as ladder logic (LD) and function block diagrams (FBD). This expansion will be crucial for the development of the Virtual PLC Auto-configurator framework, depicted in Fig.~\ref{virtualplc}, which seeks to streamline the generation of virtual PLC environments for research and deployment.

\section{SCADA Configuration} \label{sec:sgml4}
\begin{center}
 \begin{tabular}{||p{4cm}||p{10cm}||} 
 \hline\hline
 Aim & SCADA Configuration\\ 
 \hline\hline
Software & ScadaBR~\cite{scadabr} \\ 
 \hline\hline
 Outcome & SCADA Configurator Tool \\
 \hline\hline
\end{tabular}
\end{center}

The SG-ML tool provides a streamlined approach to modeling the SCADA configurator. At its core, the process relies on an XML schema and an XML file to define the SCADA system's configuration. The XML schema functions as a blueprint, specifying the structure and constraints for the address and attributes of information received from a PLC. This ensures that the configuration data is consistent and correctly formatted.

\begin{figure}[h]
\centering
{\includegraphics[width=\linewidth]{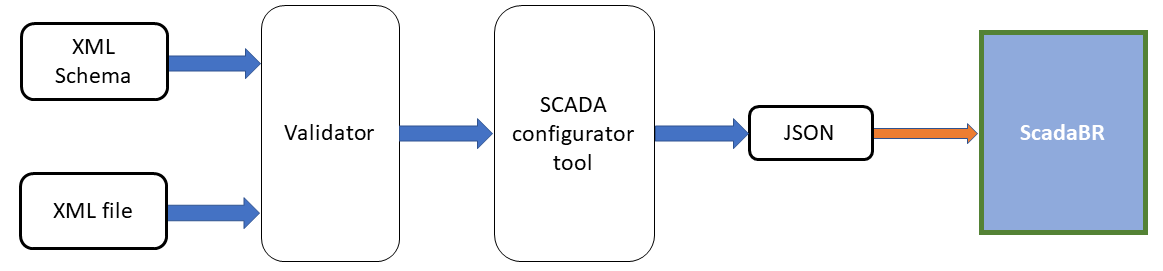}}
\caption{Framework for SCADA configurator.}
\label{scada}
\end{figure}

The first step of this automated workflow involves a validator, which checks the XML file against the schema. This validation process is critical for preventing errors and ensuring data integrity before the configuration is generated. If the XML file is successfully validated, it proceeds to the SCADA Tool Configurator. Within the configurator, the validated XML is converted into a JSON format. This conversion is often necessary as JSON is a common data interchange format for modern web-based and API-driven applications like ScadaBR. The resulting JSON file, which contains the complete and validated SCADA configuration, is then uploaded to ScadaBR for processing. This automated method significantly reduces the potential for manual configuration errors and accelerates the deployment of SCADA systems. The full end-to-end process is visually represented in the flowchart shown in Fig.~\ref{scada}.

\begin{figure}[!b]
\centering
{\includegraphics[width=\linewidth]{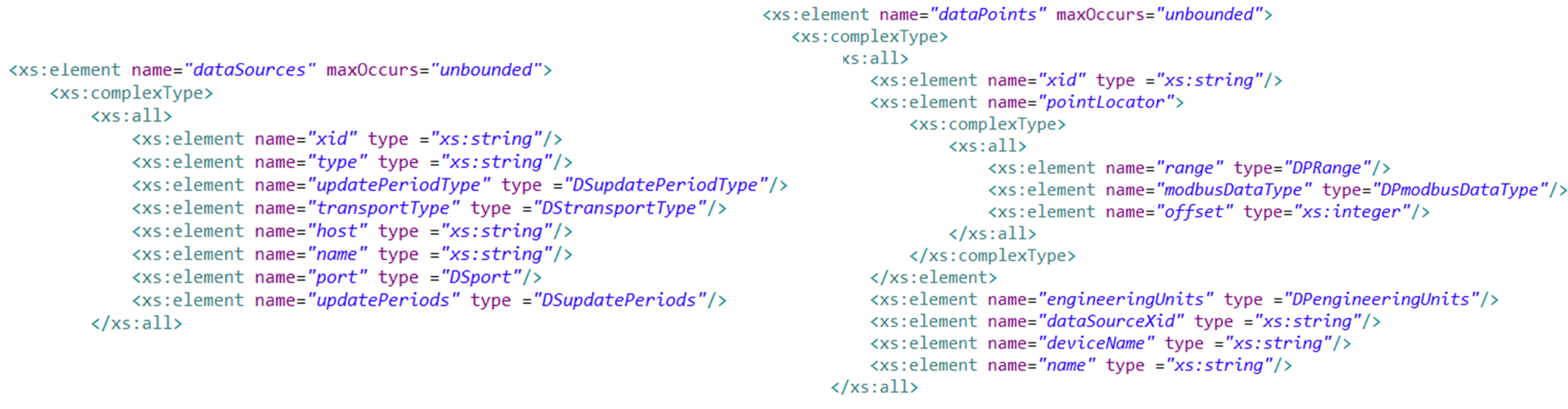}}
\caption{Schema for DataSources and DataPoints.}
\label{schdsdp}
\end{figure}

\begin{figure}[!b]
\centering
{\includegraphics[width=\linewidth]{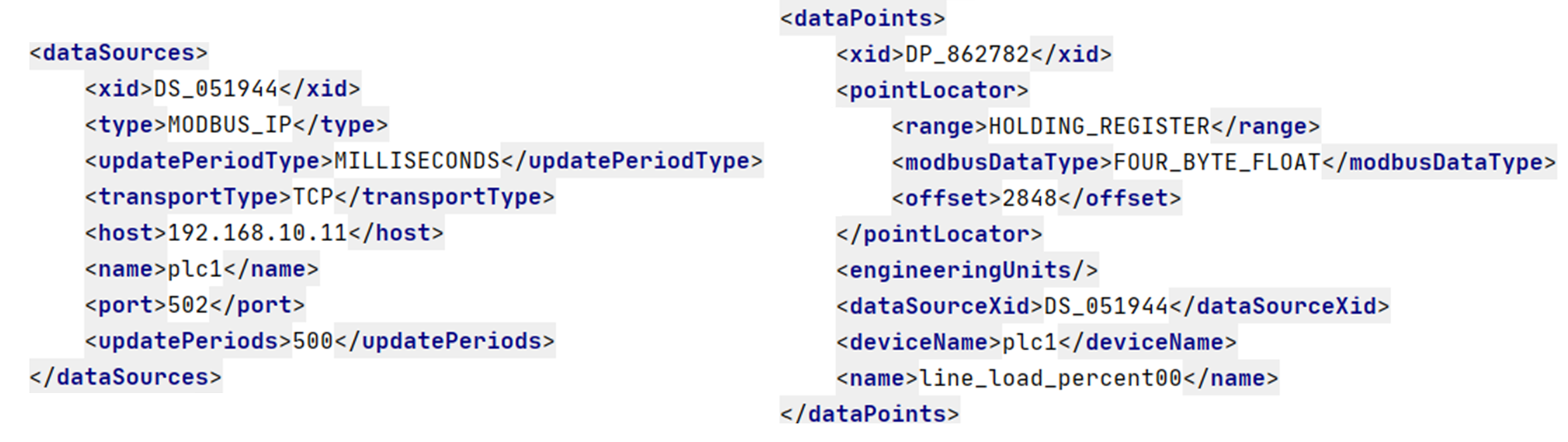}}
\caption{XML file containing DataSources and DataPoints.}
\label{xmldsdp}
\end{figure}

\begin{figure}[!h]
\centering
{\includegraphics[width=0.8\linewidth, height=5cm]{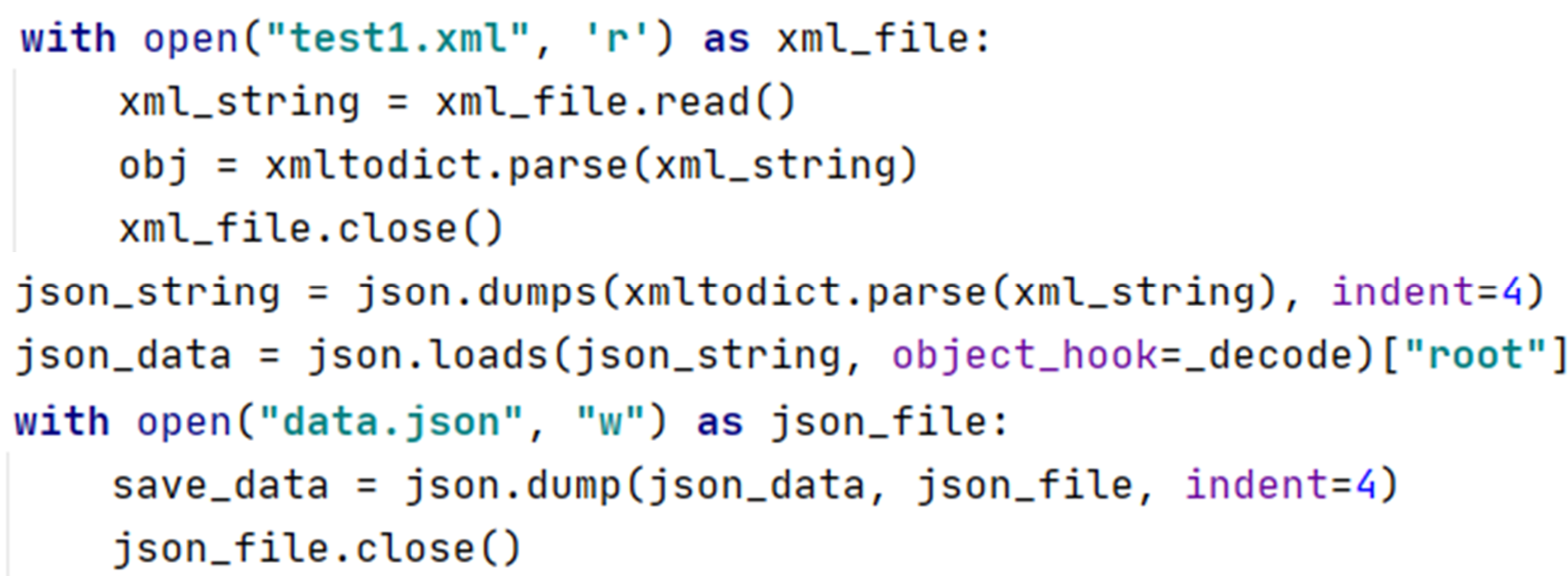}}
\caption{SCADA Tool Configurator.}
\label{scadatc}
\end{figure}

\begin{figure}[h]
\centering
{\includegraphics{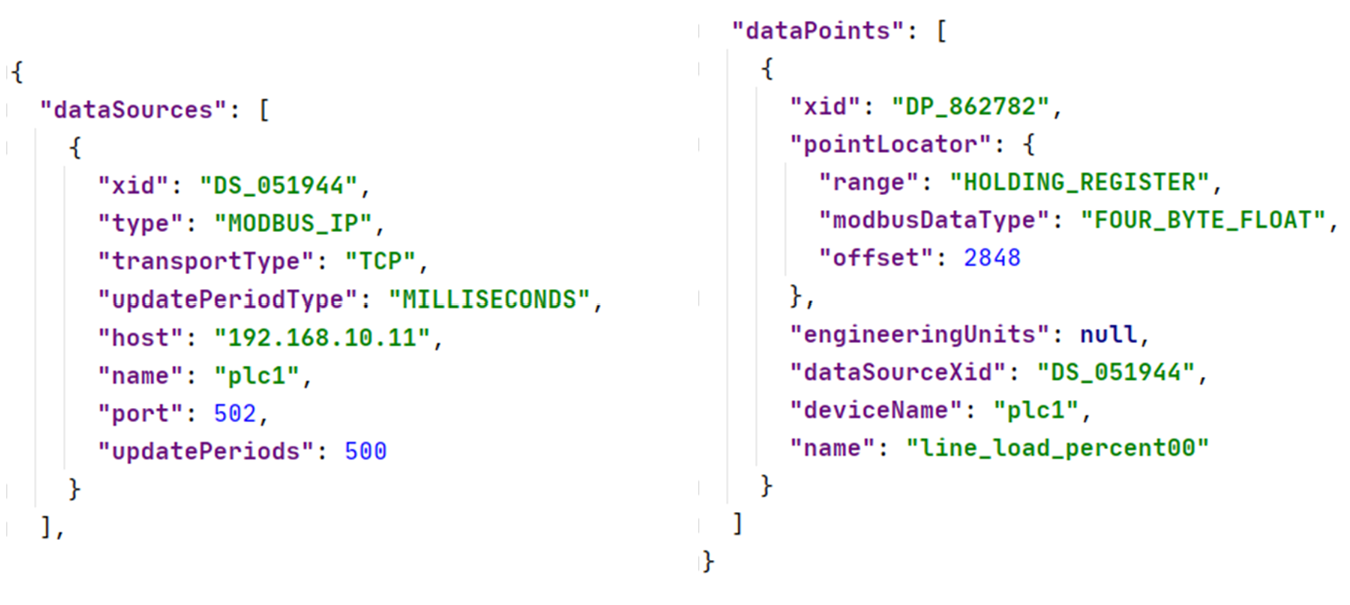}}
\caption{JSON output.}
\label{json}
\end{figure}

The crucial attributes for defining the data structure, such as "dataSources" and "dataPoints" are meticulously defined within the schema file. Figs.~\ref{schdsdp} through~\ref{json} provide a visual breakdown of this entire data transformation process:
\begin{enumerate}
    \item Fig.~\ref{schdsdp} illustrates the comprehensive schema that defines the structure for DataSources and DataPoints.
    \item Fig.~\ref{xmldsdp} provides an example of the XML file, showing how the DataSources and DataPoints are populated with specific information.
    \item Fig.~\ref{scadatc} details the inner workings of the SCADA Tool Configurator, highlighting its role in transforming the data.
    \item Fig.~\ref{json} shows the final JSON output from the configurator tool, which is ready for consumption by the ScadaBR system.
\end{enumerate}

A detailed description of all attributes included in the schema and XML files is provided in Table~\ref{tscada}.

\begin{table}[H]
\centering
\caption{Attributes in XML Proprietary Settings file}
    \begin{tabular}{|p{4cm}|p{10cm}|}
    \hline
        Attributes & Description \\
        \hline
        dataSources & Connection set up to a database from a server, e.g. Customer data base \\
        dataPoint & A piece of data that is collected during monitoring, e.g. Customer ID \\
        xid & ID of a data source or data point\\
        type & Type of data source\\
        updatePeriodType & Update period unit of data source in milliseconds, seconds etc.\\
        transportType & Type of protocol which the data source is using\\
        host & IP address of the data source\\
        name & Name of the data source or data point\\
        port & Port number of data source\\
        updatePeriods & Updating frequency of data source\\
        range & Memory range of data points\\
        offset & To offset the Modbus initial register value in data point\\
        modbusDataType & Type of memory stored for data points\\
        engineeringUnits & Measuring units of a data point\\
        dataSourceXid & Data source ID which the data points are connected to\\
        deviceName & Name of the data source which the data points are connect to\\

        \hline
    \end{tabular}
    \label{tscada}
\end{table}

\section{Multi-Substation configuration} \label{sec:sgml5}
\subsection{Multi-substation SCD file generation}

The parser combines the different SCD files of each substation to create a merged SCD file for multi-substation system. The merged SCD file for multi-substation is the concatenation of different `SubNetwork' sections. Each `SubNetwork' section contains a `Communication' section which is extracted from the SCD file of each substation. The information of all switches (i.e., `IED' elements) corresponding to different substations are listed in the merged SCD file for multi-substation. Fig.~\ref{SCD} below depicts the framework for combining different SCD files to one merged SCD file for multi-substation system. This merged SCD file is used to emulate the cyber-network topology of the multi-substation system.

\begin{figure}[!b]
\centering
{\includegraphics[width=\linewidth, height=10cm]{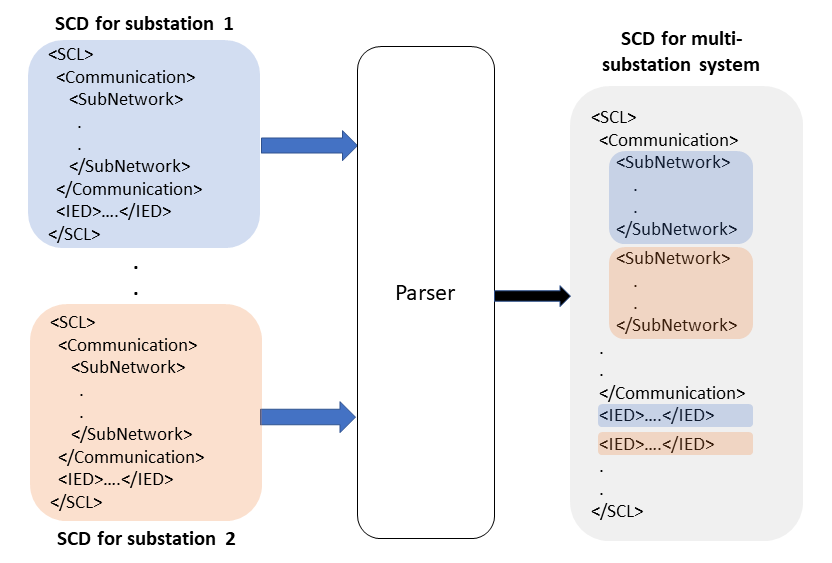}}
\caption{Framework for combining SCD files of different substations.}
\label{SCD}
\end{figure}

\subsection{Multi-substation topology description (SSD file)}
\begin{figure}[h]
\centering
{\includegraphics[width=.7\linewidth, height=6cm]{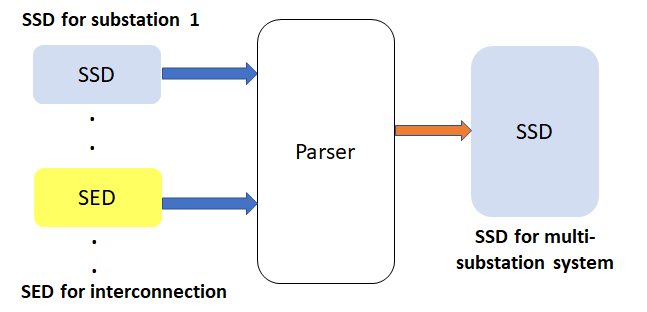}}
\caption{Framework for combining different SSD and SED files corresponding to different substations and interconnections.}
\label{SSD}
\end{figure}

The SSD files contain the description of physical electrical topology of each substation. The description of interconnection between two substations is defined by SED file. A parser tool is developed to create one single merged SSD file for multi-substation system by combining all the SSD files and SED files related to the multi-substation model. Fig.~\ref{SSD} depicts the framework for combining different SSD and SED files corresponding to different substations and interconnections respectively, to one final merged SSD file for multi-substation system.

\begin{figure}[h]
\centering
{\includegraphics[width=\linewidth, height=10.5cm]{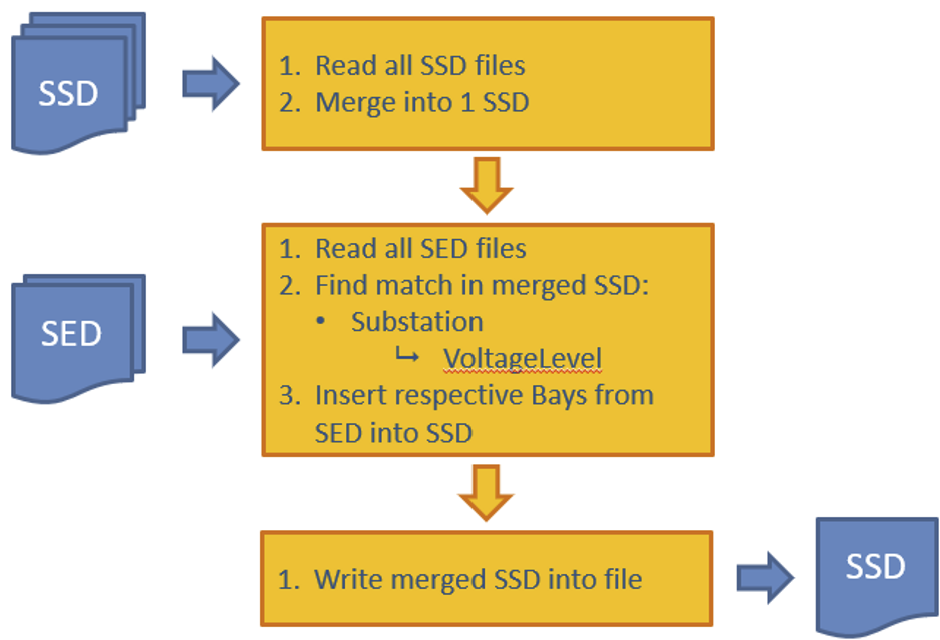}}
\caption{Process for merging SSD and SED files.}
\label{SSD_p}
\end{figure}

The SSD file typically consists of the `Substation' section, `VoltageLevel' and `Bay' sub-sections. The parser concatenates `Substation' section data from different SSD files to one final SSD file. The `VoltageLevel' tag in SED file is identified and its contents (i.e. `Bay' sub-sections) are copied under the similar `VoltageLevel' tag in final SSD file. This process is repeated for all the SED files. The procedure of creating the merged SSD file is illustrated in Fig.~\ref{SSD_p}. Thus, the final SSD file containing the complete electrical topology of multi-substation system is created.

\begin{figure}[t]
\centering
{\includegraphics[width=\linewidth]{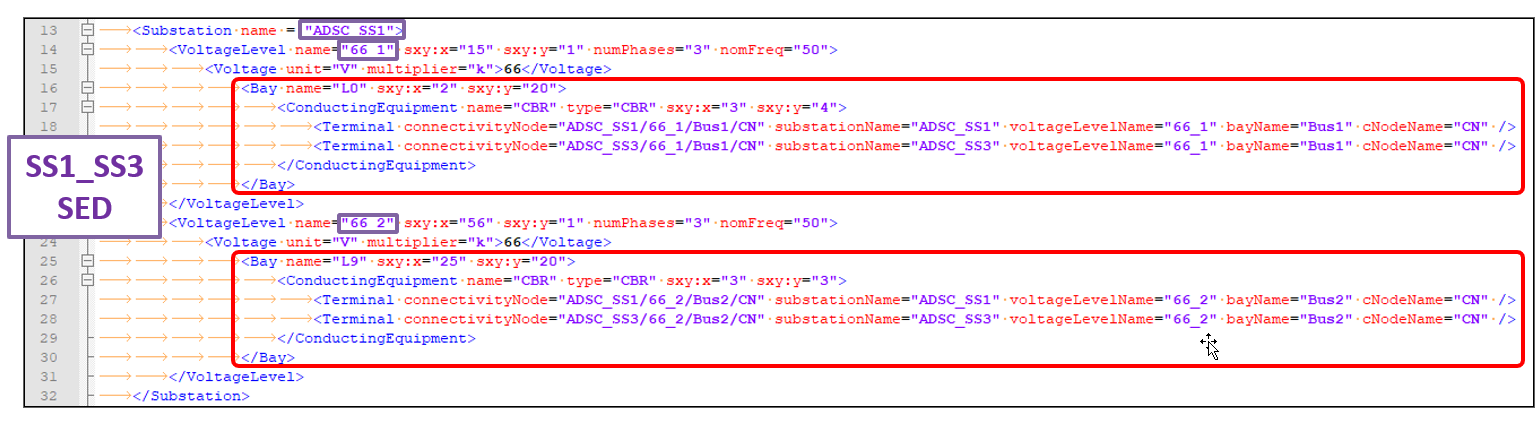}}
\caption{Sample SED file.}
\label{SED}
\end{figure}

\begin{figure}[!t]
\centering
{\includegraphics[width=\linewidth]{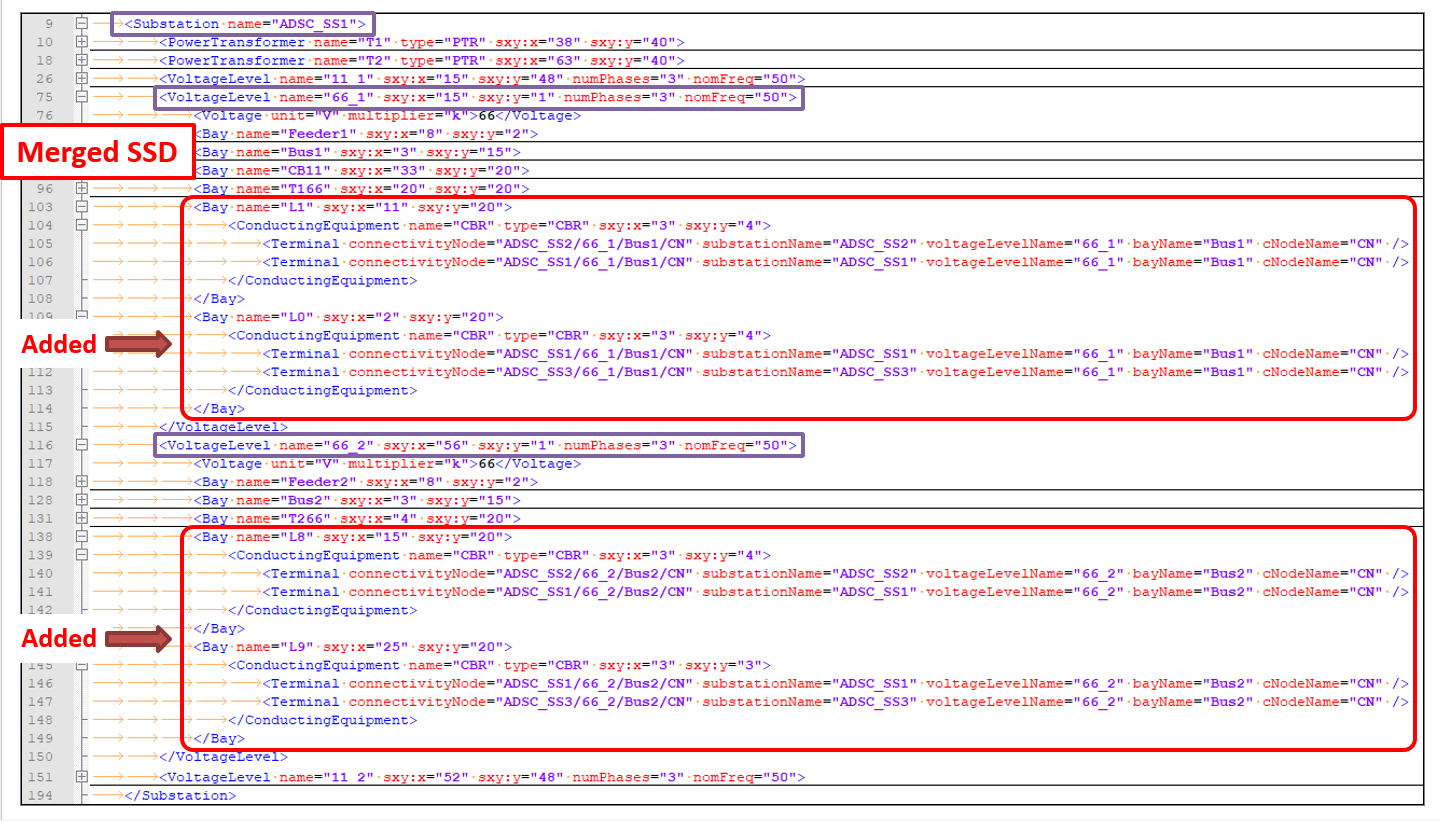}}
\caption{Sample final SSD file.}
\label{SSD_file}
\end{figure}

Fig.~\ref{SED} depicts a sample SED file containing information regarding interconnection between two substations namely ADSC\_SS1 and ADSC\_SS3. The interconnecting link is listed under `VoltageLevel' 66\_1 section and its corresponding `Bay' section. Fig.~\ref{SSD_file} shows the final SSD file generated by parser which includes the interconnection information from SED file under the `VoltageLevel' 66\_1 section.

\section{Conclusion} \label{sec:conclusion}
The SG-ML modelling language provides a comprehensive and detailed methodology for the automated generation of smart grid cyber ranges. By leveraging existing standardized models like IEC 61850 and IEC 61131, the modelling language allows for the efficient and repeatable creation of complex cyber-physical test environments. This report demonstrates the application of this modelling language to model various systems, including the EPIC testbed, a sub-transmission substation, and a three-substation network. It also outlines the use of specific Substation Configuration Language files, such as SSD and ICD, to define both the power system and cyber network configurations. The document's detailed descriptions of communication control blocks (Report and GOOSE) further underscores its capability to model realistic and dynamic interactions between IEDs and other components. 

The key contributions of this work are threefold: (1) the introduction of SG-ML as a unified modelling language for smart grid cyber ranges; (2) the integration of IEC-based standards with proprietary extensions to capture scenario-specific details; and (3) the demonstration of SG-ML across single-substation, multi-substation, and testbed environments.

\section{Related Works \& Repository}

The preliminary work was published in the industry track of Dependable Systems and Networks conference~\cite{mashima2023towards}, and the complete version appeared in IEEE Open Journal of Industrial Electronics Society~\cite{11145746}. The cloud-based implementation of Auto-SGCR using Docker containers is demonstrated in~\cite{chng2024craas}.

The associated toolchain has been open-sourced and is publicly available online at
\url{https://github.com/smartgridadsc/CyberRange}, providing access to the full software suite, example models, and documentation necessary for reproducing the experiments and deploying customized smart grid cyber ranges.

\section{Acknowledgement}
This research is supported in part by the National Research Foundation, Singapore, Singapore University of Technology and Design under its National Satellite of Excellence in Design Science and Technology for Secure Critical Infrastructure Grant (NSoE\_DeST-SCI2019-0005), and in part 
by the National Research Foundation, Prime Minister’s Office, Singapore under its Campus for Research Excellence and Technological Enterprise (CREATE) programme.

\bibliographystyle{plainnat}
\bibliography{references}
\end{document}